\documentclass[11pt]{article}

\usepackage{amssymb,amsmath,amsfonts,bbm}
\usepackage{undertilde}
\usepackage[totalwidth=17cm,totalheight=23.8cm]{geometry}

\usepackage{hyperref}

\newcommand{\im}{{\rm i }}

\newcommand{\be}{\begin{equation}}
\newcommand{\ee}{\end{equation}}
\newcommand{\beq}{\begin{eqnarray}}
\newcommand{\eeq}{\end{eqnarray}}
\newcommand{\bea}[2]{\be\label{#2}\begin{array}{#1}}
\newcommand{\eea}{\end{array}\ee}

\def\det{\,{\rm det}\, }

\def\Im{\,{\rm Im}\, }
\def\Re{\,{\rm Re}\, }

\def\({\left(}
\def\){\right)}
\def\[{\left[}
\def\]{\right]}
\def\p{\partial}
\newcommand{\de}{\mathrm{d}}
\newcommand{\I}{\mathrm{i}}

\def\11{1\!\! 1}

\def\hf{\frac{1}{2}}

\def\eps{\varepsilon}

   \def\CC {{\cal C}}
   \def\CD {{\cal D}}

   \def\CG {{\cal G}}
   \def\CH {{\cal H}}

   \def\CK {{\cal K}}

   \def\CR {{\cal R}}

   \def\CV {{\cal V}}
   
   \def\CX {{\cal X}}
   \def\CY {{\cal Y}}

\newcommand{\tH}{\lefteqn{\smash{\mathop{\vphantom{<}}\limits^{\;\sim}}}H}
\newcommand{\tE}{\lefteqn{\smash{\mathop{\vphantom{<}}\limits^{\;\sim}}}E}

\newcommand{\teps}{\tilde{\eps}}

\newcommand{\epst}{\lefteqn{\smash{\mathop{\vphantom{\Bigl(}}\limits_{\!\scriptstyle{\sim}}\atop \ }}\eps}

\newcommand{\Ht}{\lefteqn{\smash{\mathop{\vphantom{\Bigl(}}\limits_{\sim}\atop \ }}H}
\newcommand{\Et}{\lefteqn{\smash{\mathop{\vphantom{\Bigl(}}\limits_{\sim}\atop \ }}E}
\newcommand{\Nt}{\lefteqn{\smash{\mathop{\vphantom{\Bigl(}}\limits_{\sim}\atop \ }}N}
\newcommand{\Mt}{\lefteqn{\smash{\mathop{\vphantom{\Bigl(}}\limits_{\,\sim}\atop \ }}M}
\newcommand{\Gt}{\lefteqn{\smash{\mathop{\vphantom{\Bigl(}}\limits_{\sim}\atop \ }}G}

\newcommand{\bt}{\lefteqn{\smash{\mathop{\vphantom{\Bigl(}}\limits_{\!\scriptstyle{\sim}}\atop \ }}b}

\def\tphi{\tilde\phi}
\def\tpsi{\tilde\psi}

\def\Csc{\mathfrak{C}}

\def\hCG{\hat\CG}
\def\hCC{\hat\CC}
\def\hCD{\hat\CD}
\def\hCH{\hat\CH}
\def\hCY{\hat\CY}
\def\hT{\hat T}

\def\vN{\vec{N}}
\def\vM{\vec{M}}
\def\vG{\vec{G}}
\def\vF{\vec{F}}

\let\a=\alpha     \let\d=\delta
   \let\l=\lambda
\let\m=\mu    \let\n=\nu          \let\r=\rho 
  
 \let\eps=\epsilon

\newcommand{\f}{\frac}
\newcommand{\Ref}[1]{(\ref{#1})}

\newcommand{\w}{\wedge}

\begin{document}

\title{Chiral description of ghost-free massive gravity}

\vspace{0.7cm}

\author{Sergei Alexandrov$^{a,b}$, Kirill Krasnov$^c$, and Simone Speziale$^d$ }

\date{}

\maketitle

\vspace{-1cm}
\begin{center}
$^a${\it Universit\'e Montpellier 2, Laboratoire Charles Coulomb UMR 5221, F-34095,
Montpellier, France}\\
$^b${\it CNRS, Laboratoire Charles Coulomb UMR 5221, F-34095,
Montpellier, France}\\
$^c${\it Mathematical Sciences, University of Nottingham, Nottingham, NG7 2RD, UK} \\
$^d${\it \small{Centre de Physique Th\'{e}orique, CNRS-UMR 7332, Luminy Case 907, 13288 Marseille, France}}
\end{center}
\vspace{0.1cm}
\begin{abstract}
We propose and study a new first order version of the ghost-free massive gravity.
Instead of metrics or tetrads, it uses a connection together with Plebanski's chiral 2-forms as fundamental variables,
rendering the phase space structure similar to that of $SU(2)$ gauge theories.
The chiral description simplifies computations of the constraint algebra, and allows us
to perform the complete canonical analysis of the system.
In particular, we explicitly compute the secondary constraint and carry out the stabilization procedure,
thus proving that in general the theory propagates 7 degrees of freedom, consistently with previous claims.
Finally, we point out that the description in terms of 2-forms opens the door
to an infinite class of ghost-free massive bi-gravity actions.
Our results apply directly to Euclidean signature. The reality
conditions to be imposed in the Lorentzian signature appear to be more
complicated than in the usual gravity case and are left as an open
issue.
\end{abstract}

\tableofcontents

\section{Introduction}

A deceptively simple way to modify general relativity in the infrared is to give the graviton a small mass.
Massive gravity is a subject with a long history, and a most recent wave of interest
comes from a possibility that it may provide a technically natural solution to the cosmological constant problem,
see e.g. \cite{Hinterbichler:2011tt} and references therein. The theory is, however,
plagued by many difficulties. One of them is that generical non-linear completions of the Fierz-Pauli
linear theory propagate an additional scalar degree of freedom with various pathologies,
the so called Boulware-Deser ghost \cite{Boulware:1973my}. Perturbatively, the ghost manifests
itself as a scalar field with a wrong sign kinetic term. Despite this, the ghost was argued
\cite{Deffayet:2005ys} to play a role in the Vainshtein mechanism \cite{Vainshtein:1972sx} of
alleviating the so-called van Dam-Veltman-Zakharov discontinuity \cite{vanDam:1970vg}.
A related difficulty with massive gravity is that the theory becomes strongly
coupled at a very low energy scales \cite{ArkaniHamed:2002sp}.

Recently de Rham, Gabadadze and Tolley \cite{deRham:2010ik,deRham:2010kj,deRham:2011rn}
suggested that a certain non-linear completion of the Fierz-Pauli theory is free
from the ghost degree of freedom. Moreover, it was argued that the Vainshtein mechanism
is still at play, and that the cut-off is raised to a much larger energy scale,
so that the ghost free massive gravity can in particular be trusted in
the Vainshtein mechanism region.\footnote{See, however,
a recent paper \cite{Burrage:2012ja} for a reevaluation of the strong coupling issue.}
A non-perturbative formulation of the action was given in
\cite{Hassan:2011hr,Hassan:2011tf,Hassan:2011zd,Hassan:2011ea},
including its natural description as a sector of a bi-metric theory of gravity,
with the action being the sum of two Einstein-Hilbert actions for the two metrics, $g_{\mu\nu}^{+}$ and $g_{\mu\nu}^-$,
plus a local interaction $V(g^{+\m\r}g^-_{\r\n})$ between them. Linear combinations of the metric perturbations
describe a massless and a massive spin-2 particle.
Consequently, a number of papers have studied the proposed interaction
\cite{Alberte:2010qb,Chamseddine:2011mu,Kluson:2011qe,Comelli:2011wq,Golovnev:2011aa,Kluson:2011rt,
Buchbinder:2012wb,Kluson:2012gz,Kluson:2012wf,Hassan:2012wt,Hassan:2012wr,Kluson:2012zz}.
One of the key goals of these works was to establish the ghost-freeness non-perturbatively, by means of canonical analysis.
However, the latter turns out to be rather cumbersome, both in the single massive gravity and
in the bi-gravity frameworks, owing to the complicated expression of the potentials $V$ argued
to give a ghost-free theory. In particular, the potentials include square roots of metrics
and did not seem to be natural from any geometric point of view.

A new insight came from the work \cite{Hinterbichler:2012cn} (see also \cite{Chamseddine:2011mu,Deffayet:2012nr}) where
it was pointed out that these potentials take a much simpler form in the tetrad formalism.
The tetrad is a  set of four one-forms  $e^I=e^I_\mu\de x^\mu$, related to the metric by
\be
g_{\mu\nu} = e^I_\mu e^J_\nu \eta_{IJ}.
\ee
The bi-gravity ghost-free interactions, rewritten in terms of the tetrads $e_+^I$ and $e_-^I$
for the two metrics, reduce to the following 4-forms
\be\label{list}
\eps_{IJKL} e_+^I\wedge e_+^J \wedge e_+^K \wedge e_-^L,
\qquad
\eps_{IJKL} e_+^I\wedge e_+^J \wedge e_-^K \wedge e_-^L,
\qquad \eps_{IJKL} e_+^I\wedge e_-^J \wedge e_-^K \wedge e_-^L.
\ee
This is an important advance, which in particular shows that the ghost-free theory is not an artificial monstrosity,
but a rather natural theory, once the correct set of variables is identified.
The actual equivalence between this formulation and the original one in terms of
metric variables is not exact, but depends on the vanishing of a certain antisymmetric
contraction of the tetrads. This was shown in \cite{Hinterbichler:2012cn} to happen always
at the linearized level, where the condition amounts to a gauge fixing.
However, it is not true in general. Conditions for its validity were recently
spelled out in \cite{Deffayet:2012zc}.

Modulo the above discrepancy, the tetrad (or more generally vielbein) formulation also makes it completely
transparent how the primary constraint responsible for removing the ghost degree of freedom arises.
Indeed, as pointed out in \cite{Hinterbichler:2012cn}, the interaction terms (\ref{list})
are all linear in the lapse and shift functions of both metrics (due to the wedge product
used in constructing them). Varying with respect to these lapse and shift functions, one
gets the set of primary constraints. A part of this set is expected to be first class,
and realize the algebra of diffeomorphisms of the bi-metric theory. Another part is expected to
be second class. In particular, the constraints that are supposedly removing the ghost degree of
freedom are a second class pair, with one of these constraints arising as primary and one as a conjugate secondary constraint.
Unfortunately, the analysis quickly becomes rather messy, even in the simpler vielbein formulation of \cite{Hinterbichler:2012cn}.
In the metric formalism, the (difficult) computation of the secondary constraint
was carried out only recently in \cite{Hassan:2011ea} and in the St\"uckelberg formulation in \cite{Hassan:2012qv}.
The first of these papers also gives an argument as to why the stabilization procedure of requiring the preservation
of the secondary constraint by the time evolution just
fixes one of the Lagrange multipliers and does not produce any new constraints.

However, to the best of our knowledge, the complete Hamiltonian analysis has never been carried
out in the first order formalism. Given that this is the formalism where the primary
second-class constraints are immediate to see, one could possibly  expect also some
simplifications in the structure of the elusive secondary constraint.
The aims of the present paper are to carry out such first order formulation analysis,
for a model closely related to the one in \cite{Hinterbichler:2012cn},
as well as to put the ghost-free massive gravity in the framework of the so-called Plebanski chiral formalism.

Thus, in this paper we study a version of
the ghost-free massive gravity using what can be called a ``chiral" description,
based on the chiral splitting of the local Lorentz gauge group.
Such a chiral description is well-known in the case of pure gravity
(with or without cosmological constant), and goes under the name of Plebanski formulation
 of general relativity \cite{Plebanski:1977zz,Capovilla:1991qb,Krasnov:2009pu}. It can be seen as
a generalization of the tetrad formalism, where the fundamental fields are
an $SU(2)$ connection $A$ and a two-form $B$ valued in the $\mathfrak{su}(2)$ chiral subalgebra of the Lorentz algebra.
The $SU(2)$ connection replaces the spin connection, and $B$ replaces
the tetrad one-form of the tetrad description.
The metric is then constructed via the so-called Urbantke formula \cite{Urbantke},
\be\label{Urbantke}
\sqrt{-g} g_{\mu\nu} =
\f1{12} \eps_{ijk} \tilde{\eps}^{\alpha\beta\gamma\delta} B^i_{\mu\alpha} B^j_{\nu\beta} B^k_{\gamma\delta},
\ee
where the epsilon symbol with an overtilde stands for the densitized anti-symmetric tensor, that does
not need a metric for its definition. Much like the tetrad is a ``square root" of the metric,
the two-form here is its ``cubic root".
This apparent complication brings in several advantages.
First, the phase space of general relativity can now be described in the same way as an $SU(2)$ gauge theory,
with the ``triad" field for the spatial metric being canonically conjugate to the spatial part of
the $SU(2)$ gauge field. This simplifies the canonical analysis tremendously. Calculations can be performed
very efficiently, and one is able to work out the complete constraint algebra with much less effort
than in the metric (or vielbein) formulation.
Secondly, it allows to automatically take into account the extension of the theory
to a first order framework, where the connection is an independent field,
like in the Palatini formulation of general relativity.

The theory we consider in this paper is given by two copies of chiral Plebanski actions,
coupled together in essentially the same way as was advocated in \cite{Hinterbichler:2012cn}
in the context of the vielbein formulation. More precisely, the interaction we consider is
induced by the term $\d_{ij} B^{+i}\w B^{-j}$, where $B^+$ and $B^-$ are the two-form fields
for the two copies of Plebanski gravity.
We study the canonical structure of the theory, and in particular,
we explicitly compute the algebra of primary constraints and the resulting secondary constraint,
and show that the system has seven degrees of freedom so that it is free of the Boulware-Deser ghost.
We also prove that the stabilization procedure closes (under some assumptions to be clarified below).
Thus, the results of this paper provide an independent proof of the absence of the ghost and moreover
generalize it to the first order formalism.
As an illustration of the power of our formalism,
we use it to compute the Poisson brackets of two (complete, with all second-class constraints
taken into account) smeared Hamiltonians. This is a useful exercise, as it computes the quantity
that is known to have the interpretation of the physical spatial metric.
As far as we are aware,
such a computation has never been performed in any other formalism.

The price to pay for the advantages of the chiral formulation is that, in the Lorentzian signature, one has to work
with complex valued two-forms, and at some point supplement the theory with appropriate reality conditions.
For the chiral description of pure general relativity these reality conditions are well-understood.
However, new issues arise after two copies of general relativity are coupled together, and for the interacting
theory the problem of reality conditions has to be solved essentially from scratch.
This is not attempted in the present work, see, however, some comments on this issue in the main text.
In particular, an interesting observation is that the problem of finding suitable reality conditions seems to be related
to the discrepancy between tetrad and metric formulations mentioned above.
With this question lying open, our results and proofs apply to Euclidean signature bi-metric gravity.

As will be shown below, the interaction term we consider corresponds to the second term in \Ref{list},
whereas the first and third ones cannot be obtained in our formalism.
On the other hand, the use of
the 2-form chiral formalism opens a door to
an \emph{infinite} class of interactions leading to ghost-free bi-gravity theories,
as opposed to the only three terms of \Ref{list}. This comes from the possibility to modify
the dynamics described by the Plebanski action without introducing extra degrees of freedom,
see e.g. \cite{Krasnov:2008fm} for a reference. As we will argue at the end of the paper,
such modifications can be incorporated in the coupled system and are also free from the Boulware-Deser ghost.
While we will not study these theories here in any details, it is interesting to note their existence.

The organization of this paper is as follows.
In the next section we introduce the chiral Plebanski formulation of general relativity and
demonstrate how simple is its canonical analysis.
In section \ref{sec-massive}, we define the theory to be analyzed and discuss its relation to the metric formalism.
Section \ref{sec-canan} is the main part of the paper where we present
the canonical analysis of the interacting system.
The resulting canonical structure is summarized in subsection \ref{subsec-sum},
whereas in subsection \ref{subsec-mod} we discuss an infinite parameter generalization
which appears to preserve this structure. The final section \ref{sec-concl} is devoted to conclusions.
A number of appendices present some more technical results, and in particular
the computation of the secondary constraint and the Poisson bracket of two smeared Hamiltonians.

\section{Summary of the chiral Plebanski formulation}
\label{sec-review}

\subsection{Plebanski formulation of general relativity}
\label{subsec-reviewLagr}

The chiral Plebanski formulation of general relativity is described by the following action \cite{Plebanski:1977zz}
\be
S[B,A,\Psi]= \int \[B^i\wedge F^i - \frac{1}{2} (\Psi^{ij} - \lambda \delta^{ij}) B^i\wedge B^j\].
\label{Plebact}
\ee
The indices\footnote{Since the ${\mathfrak{su}}(2)$ indices are raised and lowered with the unit metric, we do not strictly
follow the rule that the indices one sums over should be in opposite positions. All repeated indices are assumed
to be summed.} $i,j=1,2,3$ are
${\mathfrak{su}}_{\mathbb C}(2)\sim {\mathfrak{so}}(3,{\mathbb C})\sim{\mathfrak{sl}}(2,{\mathbb C})$
Lie algebra ones, with the Lie algebra viewed as a vector space of complex dimension 3.
The field $A^i$ is the connection one form, $F^i=\de A^i+\hf\,\epsilon^{ijk} A^j\wedge A^k$ is its curvature,
and the field $B^i$ is a Lie algebra-valued two-form.
$\lambda$ is a multiple of the cosmological constant, and the Lagrange multiplier field $\Psi^{ij}$
is required to be traceless. Its variation enforces the so-called metricity (or simplicity) constraints
\be
B^i\wedge B^j=\frac{1}{3}\,\delta^{ij}\,B^k\wedge B_k.
\label{simpl-constr}
\ee
A solution to these constraints can be conveniently written using tetrad one-forms valued
in ${\mathfrak{so}}(4,{\mathbb C})$ as
\be\label{Bsimple}
B^i = P^{\pm\, 0i}_{IJ}\( e^I\w e^J\),
\ee
where $I=0,i$ and $P^{\pm \, KL}_{IJ}$ are two chiral projectors defined in terms of the internal Hodge star $\star$ as
\be
P^{\pm\, KL}_{IJ}:= \f12 \left({\mathbbm 1} \pm\sigma\star \right)_{IJ}{}^{KL}
= \frac{1}{4}\left( \delta_I^K \delta_J^L -\delta_I^L \delta_J^K \pm \sigma \epsilon_{IJ}{}^{KL}\right).
\ee
Here $\sigma=\im$ for the Lorentzian signature, and $\sigma=1$ for Euclidean.
Plugging the solution \eqref{Bsimple} back into the action \Ref{Plebact},
one obtains the Einstein-Cartan action plus a boundary term.
The equivalence with the Einstein-Hilbert action then follows as in
the Palatini formalism: one varies with respect to the independent connection,
obtaining an equation that identifies it in terms of the unique torsion-free Levi-Civita connection.

The action given above uses complex fields and therefore describes complexified general relativity.
To get a real theory,
one has to impose appropriate reality conditions. These are easiest to state by saying that
the metric (\ref{Urbantke}) should be a real metric of the required signature.
Such conditions can be given directly in terms of the fields that appear in the Lagrangian.
For metrics of Euclidean signature the reality condition is simply that all fields are real,
and thus the connection and the $B$ field are ${\mathfrak{so}}(3,{\mathbb R})$-valued.
For Lorentzian signature the conditions are quadratic in the fields, and can be stated as
\be\label{reality-B}
B^i\wedge (B^j)^* = 0, \qquad {\rm Re}(B^i\wedge B^i)=0.
\ee
First, one requires that the complex conjugate of any of the 3 $B^j$'s is wedge-orthogonal
to any other $B^i$ (or to itself) which gives nine real equations. In addition, one demands
the 4-form obtained as the trace of the wedge product of the $B$'s be purely imaginary.
The reason for these conditions is clear from \Ref{Bsimple}: when the $B^i$ are constructed
as the self-dual part of the two-form $e^I\wedge e^J$, with $e^I$ a real Lorentzian signature tetrad, its
complex conjugate is the anti-self-dual part, and both conditions in (\ref{reality-B})
hold.\footnote{When describing the Lorentzian theory, it is also convenient to put
an imaginary unit factor $\im$ in front of the action, as to get precisely the Einstein-Cartan action in the end.}

The reality conditions for $B^i$ also induce reality conditions on the connection $A^i$. These are obtained from
the field equation for the connection that follows from (\ref{Plebact}).
The field equation can be solved for $A^i$ in terms of $B^i$, and then the reality condition (\ref{reality-B}) implies
a reality condition on the connection. Alternatively, one can state the reality condition on $A^i$ by
passing to the Hamiltonian formulation, as shown below.

\subsection{Canonical analysis}
\label{subsec-reviewcanon}

As a preparation for a more complicated canonical analysis of a bi-gravity model,
we recall here the canonical analysis of the chiral Plebanski formulation \eqref{Plebact}.
It starts with the usual $3+1$ decomposition of the action which leads to the following Lagrangian
\be
{\cal L}=\hf\, \teps^{abc} \left(B_{0a}^i F_{bc}^i + B^i_{bc} F^i_{0a}
- (\Psi^{ij} - \lambda \delta^{ij}) B_{0a}^i B_{bc}^j \right).
\label{LagrGRdecomp}
\ee
It contains only one term with time derivatives, thus the phase space of the theory
is parametrized by the spatial components of the connection $A_a^i$, with
conjugate momentum\footnote{The tilde above the symbol keeps track of the density weight of this pseudotensor.
We will also use a tilde under the symbol to characterized negative weight pseudotensors.}
\be
\tE^{ai}:= \frac{\partial {\cal L}}{\partial (\partial_0 A_a^i)}
= \frac{1}{2}\, \teps^{abc} B_{bc}^i.
\label{defE}
\ee
The remaining fields $A_0^i, B_{0a}^i, \Psi^{ij}$ are non-dynamical. Since they enter
the Lagrangian \eqref{LagrGRdecomp} linearly, they are Lagrange multipliers
generating constraints.
To simplify the analysis, it is convenient to decompose $B_{0a}^i$ as
\be
B^i_{0a} = \(\Nt \delta^{ij} + \eps^{ijk} \bt^k + \bt^{ij}\) \tE_{a}^j,
\label{decompB}
\ee
where $\Nt, \bt^i, \bt^{ij}$ describe its trace, antisymmetric, and tracefree ($b^{ii}=0$) parts, respectively,
and we assumed the invertibility of $\tE^{ai}$ denoting by
\be
\tE^i_a = \frac{1}{2}\, \epst_{abc} \eps^{ijk} \tE^{b}_j \tE^{c}_k
\ee
its (densitized) inverse. Invertibility of the triad is a necessary condition to reproduce
general relativity, perfectly analogous to the invertibility of the metric.
After the decomposition \eqref{decompB}, the constraints obtained by varying the action
with respect to $\Psi^{ij}$ simply imply
\be\label{cansimpl}
\bt^{ij}=0.
\ee
Thus, these components of the $B_{0a}^i$ field can be omitted from now on.
In terms of the remaining fields and using the notation
\be\label{defshift}
N^a := \bt^i \tE^{a}_i,
\ee
the above Lagrangian can be rewritten as
\be
{\cal L}= \tE^{a}_i \partial_0 A_a^i + A_0^i D_a \tE^{a}_i + N^a \tE^{b}_i F^i_{ab}
+ \frac{\Nt}{2}\, \eps^{ijk} \left( \tE^{a}_i \tE^{b}_j F_{ab}^k
+ \lambda \epst_{abc} \tE^{a}_i \tE^{b}_j  \tE^{c}_k \right),
\label{decompLagrGR}
\ee
where $D_a$ is the covariant derivative defined by the connection $A_a^i$ as
$D_a X^i = \partial_a X^i + \eps^{ijk} A_a^j X^k$, and we omitted a total derivative term.
This representation identifies $\tE^a_i$ as the densitized triad determining the spatial metric,
with the last term in the Hamiltonian being the cosmological term, cubic in the triad and thus
giving the volume form of the metric.
Notice also that the covariant expression \Ref{Bsimple} is recovered from \Ref{defE} and \Ref{decompB}
upon using the constraint \Ref{cansimpl}, definition \Ref{defshift} and the standard ADM decomposition of tetrad.
In particular, $\Nt$ and $N^a$ are identified as the usual lapse and shift functions, respectively.

The resulting Lagrangian identifies the symplectic structure to be given by the Poisson bracket
\be\label{AEbrackets}
\{ A_a^i(x), \tE^b_j(y) \} = \delta_a^b \delta^i_j\, \tilde{\delta}^3(x-y).
\ee
This is the same phase space of an SU(2) gauge theory, which is the main advantage of this formulation,
and also the basis for Ashtekar's approach and loop quantum gravity \cite{Ashtekar:1991hf,Thiemann:2007zz}.
The constraints arising by varying with respect to $A_0^i, \Nt, N^a$ are the Gauss,
Hamiltonian and diffeomorphism constraints, correspondingly.
They are first class since they form a closed algebra under \Ref{AEbrackets}.
To write the algebra explicitly, we define the smeared constraints
\beq
\CG_\phi &:= &\int \de^3x \, \phi^i D_a \tE^{a}_i,
\nonumber\\
\CD_{\vN} &:=& \int \de^3x\, N^a \tE^{bi} F_{ab}^i - \CG_{N^a A_a} =
\int \de^3x\, N^a \left( \tE^{bi} \partial_a A_b^i - \partial_b (\tE^{bi} A_a^i)\right),
\\
\CH_N &:=&  \frac{1}{2} \int \de^3 x \Nt \eps^{ijk}
\left( \tE^{a}_i\tE^{b}_j F_{ab}^k
+ \lambda \epst_{abc} \tE^{a}_i \tE^{b}_j  \tE^{c}_k \right).
\nonumber
\eeq
Notice that we have shifted the initial constraint coming from $N^a$ by a term proportional to the Gauss constraint.
This is convenient because the resulting constraint precisely coincides with the generator of spatial
diffeomorphisms.
Then it is straightforward to verify the following relations
\be
\begin{split}
&
\{ \CG_{\phi_1},\CG_{\phi_2}\}=\CG_{\phi_1\times\phi_2},
\qquad\quad
\{\CD_{\vN},\CG_\phi\}=\CG_{N^a\p_a\phi},
\qquad\quad
\{\CG_\phi,\CH_{N}\}=0,
\\
&
\{\CD_{\vN},\CD_{\vM}\}=\CD_{\vec L(\vN,\vM)},
\qquad
\{\CD_{\vN},\CH_{N}\}=\CH_{L(\vN,N)},
\qquad
 \{\CH_N,\CH_M\}= \CD_{\vN(N,M)}+ \CG_{N^a(N,M) A_a^i},
\end{split}
\label{GRalg}
\ee
where we used notations
\be
\begin{split}
(a\times b)^i:=&\, \eps^{ijk} a_j b_k,
\\
L^a(\vN,\vM) :=&\,  N^b\p_b M^a-M^b\p_b N^a,
\\
L(\vN,N) := &\,  N^a\p_a \Nt-\Nt\p_a N^a,
\\
N^a(N,M):=&\, \tE^{a}_i \tE^{b}_i \(\Nt\partial_b \Mt-\Mt\partial_b \Nt\) .
\end{split}
\label{notforalgebra}
\ee
The above algebra\footnote{The constraint algebra offers a direct way of confirming the triad interpretation of $\tE^a_i$.
Indeed, as nicely explained in \cite{Hojman:1976vp}, the spatial metric can be read off
the right-hand-side of the Poisson bracket of two Hamiltonian constraints. Namely, it should appear in the function
multiplying the diffeomorphism constraint.
From the last line of \Ref{notforalgebra}, we observe that this quantity is given by $\tE^{a}_i \tE^{b}_i$,
which makes clear that the spatial metric is constructed from $\tE^a_i$ as a triad.\label{ComtwoHam}}
ensures that all constraints are preserved under the evolution generated by the Hamiltonian
\be
H_{\rm tot}=-\CG_{A_0+N^a A_a}-\CD_{\vN}-\CH_N,
\ee
which is given by a linear combination of first class constraints,
as it should be in any diffeomorphism invariant theory.

As a result, the constraints generate $3+3+1=7$ gauge symmetries which reduce $9+9=18$ dimensional phase space
to a 4-dimensional one which describes two degrees of freedom of a massless graviton.
It is worth stressing how much simpler is the above Hamiltonian analysis as compared to the usual metric ADM analysis,
or to an analogous analysis using tetrads.
It is this simplicity that we would like to take advantage of,
and write a similar ``chiral" model of the ghost-free bi-metric gravity.

Finally, we briefly discuss the reality conditions in the canonical picture.
In the case of the Euclidean signature, everything is real.
For the Lorentzian case, the spatial triad $\tE^{a}_i$ is taken to be real,
which ensures the reality of the spatial metric and can be seen to be a part of the reality
condition (\ref{reality-B}) discussed above. The remainder of the condition (\ref{reality-B})
becomes the requirement that the lapse $\Nt$ is purely imaginary and the shift $N^a$ is real.
The reality condition for the connection can then be obtained by requiring that the reality of
$\tE^{a}_i$ is preserved by the time evolution (generated by the constraints).
This fixes $A_a^i$ to be of the schematic form $A_a^i = \Gamma_a^i + \im K_a^i$,
where $\Gamma_a^i$ is the connection compatible with the triad, and $K_a^i$
is the extrinsic curvature of the spatial slice embedded in 4-geometry.
For more details on this, the reader can consult \cite{Ashtekar:1991hf}.

\section{A chiral bi-gravity model}
\label{sec-massive}

A bi-gravity theory is a model represented by two copies of general relativity coupled by a certain interaction term.
Our idea is to use the chiral description of general relativity given by the Plebanski action \eqref{Plebact}.
The simplest interaction then leads to the following action,
\be
\begin{split}
S[B^\pm,A^\pm,\Psi^\pm]=&\,\int\Bigl[ B^{+ i} \wedge F^{+ i}  + B^{- i} \wedge F^{- i}
\Bigr.\\
&\,\Bigl.
- \frac{1}{2} (\Psi^{+ ij} - \lambda^{+}\delta^{ij}) B^{+ i} \wedge B^{+ j}
- \frac{1}{2} (\Psi^{- ij} - \lambda^{-}\delta^{ij}) B^{-i}\wedge B^{-j}  + 2\alpha B^{+i} \wedge B^{-i}\Bigr],
\end{split}
\label{ourtheory}
\ee
where the indices $\pm$ distinguish the fields from the two sectors.
The coupling constant $\alpha$ will later on get related to the mass of the second graviton.

In absence of the interaction term, i.e. for $\alpha=0$,
the gauge group of the theory is given by the direct product of the symmetries of each of the copies
of Plebanski theory, which include diffeomorphisms and local gauge rotations.
The interaction breaks the total group
${\it Diff}_+\times {\rm SO}_+(3)\times {\it Diff}_-\times {\rm SO}_-(3)$
to its diagonal subgroup, and it is this fact that is responsible
for a larger number of propagating modes in \eqref{ourtheory} as compared to the non-interacting theory.

Our model can easily be reformulated in terms of the tetrads instead of the two-forms $B^\pm$.
Indeed, in each $\pm$-sector the variation with respect to the Lagrange multipliers $\Psi^{\pm ij}$ imposes
the simplicity constraints \eqref{simpl-constr}, which imply that $B^\pm$ are the self-dual
projections\footnote{In \eqref{Bsimple} the indices $\pm$ denote the chirality and should not be confused with
the indices $\pm$ of this section which distinguish the two Plebanski sectors.}
of the two-forms $e_\pm^I\wedge e_\pm^J$ as in \eqref{Bsimple}. Thus, each Plebanski term reduces to
the Einstein-Cartan action for the tetrad with its own cosmological term.
At the same time, the interaction term takes the following form
\be\label{BB}
\frac{\sigma}{2}\, \epsilon_{IJKL}  e_{+}^{I}\wedge  e_{+}^{J}\wedge  e_{-}^{K}\wedge  e_{-}^{L}
+ e_+^I\wedge  e_+^J \wedge  e_{- I}\wedge  e_{- J}.
\ee
The first contribution is exactly the symmetric interaction term in \eqref{list} which has been argued
to generate a ghost-free massive gravity \cite{Hinterbichler:2012cn}. This is why we take
\eqref{ourtheory} as our starting point. Notice also the presence of the second contribution,
 not of the type \Ref{list} considered in \cite{Hinterbichler:2012cn}.
It is consistent with all symmetries of the theory, except it is parity-odd, unlike the rest of the Lagrangian.

On the other hand, the relation of our model to the metric formulation is more subtle.
The reason for this is that the equivalence between the polynomials \Ref{list}, and the ghost-free interactions between
two metrics originally proposed in \cite{deRham:2010ik,Hassan:2011hr} and related references,
holds only if the tetrads satisfy the symmetry property
\be\label{eesym}
e_{+[\mu}^I e_{-\nu]I} = 0.
\ee
In \cite{Hinterbichler:2012cn} it was shown that this condition can always be realized
in perturbation theory, where it basically amounts to a choice of gauge.
However, as it was recently pointed out in \cite{Deffayet:2012zc}, it does not hold
true in general. Hence, already the tetrad formulation turns out to be slightly different from the metric
formulation. Our model introduces a further difference due to the presence
of the second contribution in \Ref{BB}.
In our formulation, this term simply can not be discarded, as the whole structure \Ref{BB} comes at once.
It is interesting to note that this contribution vanishes provided the condition \eqref{eesym} holds.
This fact puts our formulation of the ghost-free bi-gravity theory on the same footing as the tetrad formulation
as both of them coincide with the one originally proposed in \cite{deRham:2010ik,Hassan:2011hr}
only for the configurations satisfying \Ref{eesym}. We further comment on these issues below.

Let us also make a few comments about reality conditions.
In the case of Euclidean signature, as in the usual Plebanski theory,
all fields can be taken to be real and one finds real Euclidean bi-metric gravity.
In contrast, in Lorentzian signature, the situation with the reality conditions in the presence of
the interaction term becomes more intricate than in the single gravity case.
The simplest idea is to impose the same reality conditions which have been discussed in section \ref{subsec-reviewLagr}
in each of the two sectors. This renders each metric real and ensures also
that each of the Plebanski actions is real modulo a surface term
\cite{Alexandrov:1998cu} provided one multiplies the whole action by a factor of $\im$.
As for the interaction term \Ref{BB}, the extra $\I$
cancels the factor of $\sigma=\im$ in the first term in \eqref{BB} and makes it real as well.
However, the second term in \eqref{BB} then becomes purely imaginary, thus the total action is not real.
This implies that for Lorentzian
signature the standard reality conditions of Plebanski theory are not appropriate
and should be modified to deal with the bi-gravity case.
While it is possible that there is some more sophisticated choice, we do not consider
this issue any further in the present paper, and content ourselves to
perform the analysis for the Euclidean signature when
all the fields are real (or, equivalently, for the complexified theory, when all fields are complex).
Nevertheless, let us note that, as pointed out in the previous paragraph,
the term spoiling the reality of the action vanishes
on configurations satisfying \eqref{eesym}. As a result, we come
to an interesting observation that the condition of the reality of the action
(for the standard reality conditions on the fields) precisely coincides with the condition
for the equivalence with the metric formulation.
However, just like it is not obvious how to include that restriction from
the start in the tetrad formalism, it is also not obvious how to achieve the reality of \Ref{ourtheory}
in a simple geometric way.

\section{Canonical analysis and ghost-freeness}
\label{sec-canan}

In this section we perform a careful canonical analysis of the model \eqref{ourtheory},
obtain explicit expressions for all constraints, including the secondary constraint responsible for the absence of
the Boulware-Deser ghost, and compute their algebra.
The results of this analysis are summarized below in subsection \ref{subsec-sum}.

\subsection{$3+1$ decomposition and field redefinition}
\label{subsec-31decomp}
The Hamiltonian analysis of the action \eqref{ourtheory} can be done in the same way as in section \ref{subsec-reviewcanon}.
The first step is to perform the $3+1$ decomposition which gives rise to two actions of the form \eqref{LagrGRdecomp}
for the $\pm$-fields, plus the term coupling the two sectors
\be
{\cal L}_{int}= \alpha\left( B^{+ i}_{0a} B^{- i}_{bc} + B^{- i}_{0a} B^{+ i}_{bc} \right) \teps^{abc} .
\label{intterm}
\ee
In both sectors we can introduce the fields $\tE$, $\Nt$ and $N^a$ as before in \eqref{defE} and \eqref{decompB}.
The two traceless components $b^{\pm ij}$ are again constrained to vanish by the simplicity constraints
generated by $\Psi^{\pm ij}$.
In terms of the remaining components, the interaction term \eqref{intterm} takes the following form
\be
\begin{split}
{\cal L}_{int} =&\, \alpha \left( \Nt^+ \epst_{abc} \eps^{ijk}
\tE^{+ b}_j \tE^{+ c}_k + 2\epst_{abc} N^{+ b} \tE^{+ ci} \right) \tE^{- a}_i
\\
&\,
+  \alpha \left( \Nt^- \epst_{abc} \eps^{ijk} \tE^{- b}_j \tE^{- c}_k
+ 2\epst_{abc} N^{- b} \tE^{- ci} \right) \tE^{+ a}_i .
\end{split}
\label{inttermsplit}
\ee

The canonical form of the action is then the sum of two copies of \Ref{decompLagrGR} plus the above interaction.
Being linear in lapse and shift, the interaction term directly contributes to
the constraints already present in the $\pm$-sectors.
As will be shown below, these contributions break the ``off-diagonal" part of the gauge symmetry
of the non-interacting theory, leaving only its ``diagonal" part as the symmetry of the full theory.
For this reason it is desirable to introduce variables that will be adapted to this pattern.
Motivated by this, we do the following field redefinition:
\begin{subequations}\label{linchange}
\be
\begin{array}{rclcrcl}
\tE^{a}_i &:=& \tE^{+a}_i+\tE^{-a}_i,
&\qquad &
\tH^{a}_i &:=& \tE^{+a}_i-\tE^{-a}_i,
\\
A_a^i &=& \(A_a^{+ i} + A_a^{- i}\)/2, &
\qquad &
\eta_a^i &:=& (A_a^{+ i} - A_a^{- i})/2.
\end{array}
\label{changevar}
\ee
Of course, only the variable $A_a^i$ remains a connection, with the field $\eta_a^i$
transforming as a matter field under the diagonal ${\rm SO}(3)$.
The curvatures $F^{\pm}$ decompose as
\be\nonumber
F^{\pm i}_{ab}(A_a^{\pm i} ) = F^i_{ab}(A_a^{i} ) \pm 2 D_{[a} \eta^i_{b]} + \eps^{ijk} \eta^j_a \eta^k_b,
\ee
where $D_a$ is the covariant derivative with respect to the connection $A_a$.
Performing a similar redefinition for the Lagrange multipliers
\be
\begin{array}{rclcrcl}
\phi^i &:=& (A_0^{+ i} + A_0^{- i})/2,
& \qquad &
\psi^i &:=& \( A_0^{+ i} - A_0^{- i}\)/2,
\\
N^a &:=&  (N^{+ a} + N^{- a})/2,
& \qquad &
G^a &:=& \( N^{+ a} - N^{- a}\)/2,
\\
\Nt &:=& (\Nt^+ + \Nt^-)/4,
&\qquad &
\Gt &:=& \( \Nt^+ - \Nt^-\)/4,
\end{array}
\label{changeLagr}
\ee\end{subequations}
allows to disentangle the diagonal and off-diagonal sectors, thereby making the diagonal symmetries manifest.
Distinguishing the diagonal and off-diagonal constraints by putting a hat on the latter,
the total action after the change of variables takes the following canonical form
\be
S=\int \de t\,\de^3 x\( \tE^{ai} \partial_0 A_a^i + \tH^{ai} \partial_0 \eta_a^i
+\phi^i\CG_i+\psi^i\hCG_i+N^a\CC_a+G^a\hCC_a+\Nt\CH+\Gt\hCH\).
\label{decS}
\ee

Before we give the explicit form of the constraints, it is useful to contrast the above change
of variables with what is usually done in the metric formalism.
There one keeps the original metrics as fundamental variables, and mixes only
the Lagrange multipliers (using ``geometric averages'' rather than algebraic ones as above),
see e.g. \cite{Damour:2002ws}. However, upon linearization the mass eigenstates
turn out to be the sums and differences of the (perturbations of the) two initial metrics,
$h^+_{\m\n}\pm h^-_{\m\n}$.
Extending this redefinition to the full theory, one may work with
\be
g_{\mu\nu}=g^+_{\mu\nu}+g^-_{\mu\nu},
\qquad
q_{\mu\nu}=g^+_{\mu\nu}-g^-_{\mu\nu}.
\label{usualdec}
\ee
This is morally what we are doing here.
In particular, we expect that perturbations of $\tE^a_i$ describe the degrees of freedom
of a massless graviton, and those of $\tH^a_i$ a massive spin 2 field.
However, there is no reason to expect that the linear transformation \eqref{linchange} or \eqref{usualdec}
decouples the physical metric and the massive field beyond the linearized theory.
In fact, even without discussing the coupling with matter, the physical metric can be
identified by computing the Poisson bracket of two smeared Hamiltonian constraints,
as explained in footnote \ref{ComtwoHam}.
For our model this calculation is performed in appendix \ref{ap-twoH} and shows that the physical spatial metric
is a non-trivial function in phase space, and does not coincide neither with $g_{ab}$ defined by $\tE^a_i$,
nor with the ones defined in \eqref{usualdec}.
The same situation arises in the metric formalism, see e.g. \cite{Kluson:2012ps}.
Thus, disentangling the relation between the physical spatial metric
(as appears in the algebra of diffeomorphisms) and the phase space variables
is a complicated task, whatever formulation is used.
Despite the redefinition \Ref{linchange} does not fulfil this task, it does disentangle
the constraint algebra and therefore is very convenient.

Let us now discuss the two Hamiltonian constraints which read as
\be
\begin{split}
\CH=& \, \frac{1}{2} \eps^{ijk} (\tE^{ai}\tE^{bj}
+  \tH^{ai}\tH^{bj})(F_{ab}^k + \eps^{klm} \eta_a^l \eta^m_b )
+  \eps^{ijk} \tE^{ai} \tH^{bj} (D_a \eta_b^k - D_b \eta_a^k)
\\
& + \frac{1}{2}\, \eps^{ijk} \epst_{abc}
\left[ \(\alpha+\frac{\lambda}{2}\)\tE^{a}_i \tE^{b}_j \tE^{c}_k
+3\beta  \tE^{a}_i \tE^{b}_j \tH^{c}_k
+\(\frac{3\lambda}{2}-\alpha\)\tE^{a}_i \tH^{b}_j \tH^{c}_k
+\frac{\beta}{2}\, \tH^{a}_i \tH^{b}_j \tH^{c}_k\right]
\end{split}
\ee
and
\be
\begin{split}
\hCH= \, &
\eps^{ijk} \tE^{ai}\tH^{bj} (F^k_{ab} + \eps^{klm} \eta^l_a \eta^m_b)
+  \eps^{ijk} (\tE^{ai}\tE^{bj}+\tH^{ai}\tH^{bj}) D_a \eta_b^k
\\
&+ \frac{1}{2}\, \eps^{ijk} \epst_{abc}
\left[ \frac{\beta}{2}\, \tE^{a}_i \tE^{b}_j \tE^{c}_k
+\(\alpha+ \frac{3\lambda}{2}\) \tE^{a}_i \tE^{b}_j \tH^{c}_k
+\frac{3\beta}{2}\,\tE^{a}_i \tH^{b}_j \tH^{c}_k
+\(\frac{\lambda}{2}-\alpha\) \tH^{a}_i \tH^{b}_j \tH^{c}_k\right],
\end{split}
\ee
where we denoted
\be
\lambda = \hf\(\lambda^++\lambda^-\),
\qquad
\beta = \hf\(\lambda^+-\lambda^-\).
\ee
These expressions, and the subsequent analysis, can be significantly simplified
by restricting the parameters as
\be
\label{lambda}
\lambda=-2\alpha,
\qquad
\beta=0.
\ee
These restrictions can be justified as follows.
First, the combination $\alpha+\lambda/2$ can be identified with the effective cosmological
constant of the combined system. Since the presence of the cosmological constant
has a very little effect on the constraint algebra, we can safely set it to zero.
The second restriction arises from the observation that it removes from the Hamiltonian constraint $\CH$
the term linear in $\tH$ whose presence would imply
that the Lagrangian for the massive field contains a tadpole and is not in its canonical form.

After we have made the choice \eqref{lambda}, one remains with only one free parameter $\alpha$,
which can be identified with the mass of the $\tH$ field.
The full set of simplified constraints resulting from this choice can be found in \eqref{constraints}
and will be discussed in the next subsection.

\subsection{Primary constraints}
\label{subsec-primary}

The canonical form of the action \eqref{decS} shows that the phase space of the theory is spanned by
the fields $A,\tE,\eta,\tH$ and carries the (pre-)symplectic structure encoded in the following
canonical Poisson brackets
\be
\{ A_a^i(x), \tE^b_j(y) \} = \delta_a^b \delta^i_j \,\tilde{\delta}^3(x-y),
\qquad
\{ \eta_a^i(x), \tH^b_j(y) \} = \delta_a^b \delta^i_j\, \tilde{\delta}^3(x-y).
\ee
The other variables introduced in \eqref{changeLagr} are the Lagrange multipliers for
the primary constraints which have the following (smeared) expressions
\be
\begin{split}
\CG_\phi :=&\,  \int \de^3x \, \phi^i \[ D_a \tE^{a}_i + \eps_{ijk} \eta_a^j \tH^{a k}\],
\\
\hCG_\psi :=&\, \int \de^3 x\, \psi^i \[ D_a \tH^{a}_i + \eps_{ijk} \eta^j_a \tE^{ak} \],
\\
\CC_{\vN}:=&\,  \int \de^3x\,  N^a\[ \tE^{b}_i (F_{ab}^i + \eps^{ijk} \eta_a^j \eta_b^k)
+ \tH^{b}_i (D_a \eta_b^i - D_b \eta_a^i)\],
\\
\hCC_{\vG}:= &\, \int \de^3x \, G^a\[ \tE^{b}_i (D_a \eta_b^i - D_b \eta_a^i)
+  \tH^{b}_i (F_{ab}^i + \eps^{ijk} \eta_a^j \eta_b^k)
- 2 \alpha  \epst_{abc} \tE^{b}_i\tH^{c}_i\],
\\
\CH_N:= &\, \int \de^3 x\, \Nt\biggl[ \frac{1}{2}\,\eps^{ijk} (\tE^{a}_i\tE^{b}_j
+  \tH^{a}_i\tH^{b}_j)(F_{ab}^k + \eps^{klm} \eta_a^l \eta^m_b )
+\eps^{ijk} \tE^{a}_i \tH^{b}_j (D_a \eta_b^k - D_b \eta_a^k)
\biggr.
\\
&\, \biggl.
 - 2\alpha  \eps^{ijk} \epst_{abc} \tE^a_i \tH^{b}_j \tH^{c}_k\biggr],
\\
\hCH_G:=&\,   \int \de^3 x\, \Gt\Bigl[ \eps^{ijk} \tE^{a}_i\tH^{b}_j (F^k_{ab} + \eps^{klm} \eta^l_a \eta^m_b)
+  \eps^{ijk} (\tE^a_i\tE^{b}_j+\tH^{a}_i\tH^{b}_j) D_a \eta_b^k
\Bigr.
\\
&\, \Bigl.
- \alpha \eps^{ijk} \epst_{abc}
\left( \tE^{a}_i \tE^{b}_j \tH^{c}_k +\tH^{a}_i \tH^{b}_j \tH^{c}_k\right)\Bigr].
\end{split}
\label{constraints}
\ee
In complete analogy with the single copy of Plebanski theory,
it is convenient to shift the constraints $\CC_{\vN}$ and $\hCC_{\vG}$ by a linear combination of the two
Gauss constraints. Namely, we define
\beq
\CD_{\vN}&: =&\CC_{\vN}-\CG_{N^a A_a}-\hCG_{N^a\eta_a}=
\int d^3x \, N^a\[\tE^b_i\p_a A_b^i-\p_b\(\tE^b_i A_a^i\)
+\tH^b_i\p_a \eta_b^i-\p_b\(\tH^b_i \eta_a^i\) \],
\nonumber\\
\hCD_{\vN}&:=&\hCC_{\vN}-\CG_{N^a \eta_a}-\hCG_{N^a A_a}=
\int d^3x \, N^a\[\tE^b_i\p_a \eta_b^i-\p_b\(\tE^b_i \eta_a^i\)+\tH^b_i\p_a A_b^i-\p_b\(\tH^b_i A_a^i\)\]
\\
&& - 2\alpha \int d^3x \, N^a\[\epst_{abc}\tE^b_i\tH^c_i\].
\nonumber
\eeq

It is now a straightforward although tedious exercise to compute the algebra of the primary constraints.
In contrast to the case of pure gravity, it is not anymore first class: some of the commutators acquire
contributions which are non-vanishing on the constraint surface.
The first class part of the algebra is as follows,
\begin{subequations}\be\begin{array}{rclrclrcl}
\{ \CG_{\phi_1},\CG_{\phi_2}\}&=&\CG_{\phi_1\times\phi_2},
&\qquad
\{ \CG_{\phi},\hCG_{\psi}\}&=&\hCG_{\phi\times\psi},
&\qquad
\{ \hCG_{\psi_1},\hCG_{\psi_2}\}&=&\CG_{\psi_1\times\psi_2},
\\
\{\CD_{\vN},\CG_\phi\}&=& \CG_{N^a\p_a\phi},
&
\{\CD_{\vN},\hCG_\psi\}&=&\hCG_{N^a\p_a\psi},
&
\{\hCD_{\vG},\CG_\phi\}&=&\hCG_{G^a\p_a\phi},
\\
\{\CG_\phi,\CH_{N}\}&=&0,
&
\{\CG_\phi,\hCH_{G}\}&=&0,
&
\{\CD_{\vN},\CD_{\vM}\}&=& \CD_{\vec L(\vN,\vM)},
\\
\{\CD_{\vN},\CH_{N}\}&=&\CH_{\vec L(\vN,N)},
&
\{\CD_{\vN},\hCH_{G}\}&=&\hCH_{\vec L(\vN,G)},
&
\{\CD_{\vN},\hCD_{\vG}\}&=&\hCD_{\vec L(\vN,\vG)},
\end{array}\ee
and
\be\begin{split}
\{\CH_N,\CH_M\}\,=&\,\, \CD_{\vec V(N,M)}+\hCD_{\vec U(N,M)}
 + \CG_{V^a(N,M) A_a+U^a(N,M) \eta_a}+ \hCG_{V^a(N,M) \eta_a+U^a(N,M) A_a},
\\
 \{\hCH_G,\hCH_F\}\,=&\,\, \CD_{\vec V(G,F)}+\hCD_{\vec U(G,F)}
 +\CG_{V^a(G,F) A_a+U^a(G,F) \eta_a}+ \hCG_{V^a(G,F) \eta_a+U^a(G,F) A_a},
\end{split}
\ee
where we used the notations introduced in \eqref{notforalgebra} as well as
\be\nonumber
\begin{split}
V^a(N,M) :=&\, \(\tE^a_i\tE^b_i+\tH^a_i\tH^b_i\)\(\Nt\p_b \Mt-\Mt\p_b \Nt\),
\\
U^a(N,M) :=&\, \(\tE^a_i\tH^b_i+\tH^a_i\tE^b_i\)\(\Nt\p_b \Mt-\Mt\p_b \Nt\).
\end{split}
\ee
The constraints $\CG_\phi$ and $\CD_{\vN}$ weakly commute with all other constraints
and form exactly the same subalgebra as in \eqref{GRalg}.
It is thus clear that they represent the generators
of the usual gauge and diffeomorphism transformations.

The remaining Poisson brackets are given by
\beq
\{\hCD_{\vG},\hCG_\psi\}&=&
\CG_{G^a\p_a\psi}
+2\alpha \int d^3x\, \psi^iG^a \epst_{abc} \eps^{ijk} \(\tE^{b}_j\tE^{c}_k - \tH^{b}_j\tH^{c}_k\),
\nonumber\\
\{ \hCG_\psi, \CH_{N} \} &=&
-4\alpha \int d^3x\, \Nt\psi^i \epst_{abc}
 \tE^a_i\tE^{b}_j\tH^{c}_j,
\nonumber\\
\{ \hCG_\psi, \hCH_{G} \} &= &
-4\alpha \int d^3x\, \Gt\psi^i \epst_{abc}
 \tH^a_i\tE^{b}_j\tH^{c}_j,
\nonumber\\
\{\hCD_{\vG},\hCD_{\vF}\}&=&
\CD_{\vec L(\vG,\vF)}
+4\alpha\int d^3 x\, \epst_{abc}G^a F^b \(\tE^c_i\p_d\tE^d_i-\tH^c_i\p_d\tH^d_i\),
\nonumber\\
\{\hCD_{\vG},\CH_{N}\}&=&
\hCH_{L(\vG,N)}+\alpha\CG_{\phi_1(N,\vG)}-\alpha\hCG_{\phi_2(N,\vG)}
\label{hCD-hCH}\\
&+&
4\alpha\int d^3 x\, \Nt \epst_{abc}\[ G^d A_d^i\tE^a_i\tE^b_j\tH^c_j
+G^a\eta_d^i\(\tE^b_i\(\tE^c_j\tE^d_j-\tH^c_j\tH^d_j\)+\tH^d_i \tE^b_j\tH^c_j\)\],
\nonumber\\
\{\hCD_{\vG},\hCH_{G}\}&=&
\CH_{L(\vG,G)}+\alpha\CG_{\phi_2(G,\vG)}-\alpha\hCG_{\phi_1(G,\vG)}
\nonumber\\
&+&
4\alpha\int d^3 x\, \Gt \epst_{abc}\[ G^d A_d^i\tH^a_i\tE^b_j\tH^c_j
+G^a\eta_d^i\(\tH^b_i\(\tE^c_j\tE^d_j-\tH^c_j\tH^d_j\)+ \tE^d_i \tE^b_j\tH^c_j\)\],
\nonumber\\
\{\CH_N,\hCH_G\}&= &
-\CD_{\vec U(N,G)}-\hCD_{\vec V(N,G)}
- \CG_{U^a(N,G) A_a+V^a(N,G) \eta_a}- \hCG_{U^a(N,G) \eta_a+V^a(N,G) A_a}
\nonumber\\
&-&
2\alpha\int \de^3 x \,
\Nt\Gt \eps_{abc}\eps^{ijk}
\(\tE^a_l\tE^b_j\tE^c_k\tE^d_i+\tH^a_l\tH^b_j\tH^c_k\tH^d_i
-\tE^a_l\tH^b_j\tH^c_k\tE^d_i-\tH^a_l\tE^b_j\tE^c_k\tH^d_i\)\eta^l_d,
\nonumber\eeq
\label{fullalgebra}
\end{subequations}
where
\be\label{L-vN-N}
\begin{split}
\phi_1^i(N,\vG) :=&\,
2\Nt G^a\eps^{ijk}\epst_{abc}\tE^b_j\tH^c_k,
\\
\phi_2^i(N,\vG) :=&\,
\Nt G^a\eps^{ijk}\epst_{abc}\(\tE^b_j\tE^c_k+\tH^b_j\tH^c_k\).
\end{split}
\ee
As is expected, all contributions non-vanishing at the constraint
surface are proportional to the mass parameter $\alpha$.

\subsection{Secondary constraint}
\label{subsec-secondary}

Since the primary constraints do not form a closed algebra,
the Dirac's stabilization procedure does not stop at the first step and
we have to study the conditions ensuring that the time evolution preserves the constraints.
The evolution is generated by the total Hamiltonian given by a linear combination of the primary constraints
\be
H_{\rm tot}=-\CG_{\tphi}-\hCG_{\tpsi}-\CD_{\vN}-\hCD_{\vG}-\CH_N-\hCH_G,
\label{Htot1}
\ee
where we denoted $\tphi=\phi+N^a A_a+G^a \eta_a$ and $\tpsi=\psi+N^a\eta_a+G^a A_a$.
Since $\CG_i$ and $\CD_a$ weakly commute with all primary constraints, the requirement of their stability
under evolution does not generate any conditions.
For other constraints one finds
\be
\begin{split}
\dot\hCG_i \approx&\, \{\hCD_{\vG},\hCG_i\}+\{\CH_N,\hCG_i\}+\{\hCH_G,\hCG_i\}\approx 0,
\\
\dot\hCD_a \approx&\, \{\hCG_{\tpsi},\hCD_a\}+\{\hCD_{\vG},\hCD_a\}+\{\CH_N,\hCD_a\}+\{\hCH_G,\hCD_a\}\approx 0,
\\
\dot\CH \approx&\, \{\hCG_{\tpsi},\CH\}+\{\hCD_{\vG},\CH\}+\{\hCH_G,\CH\}\approx 0,
\\
\dot\hCH \approx&\, \{\hCG_{\tpsi},\hCH\}+\{\hCD_{\vG},\hCH\}+\{\CH_N,\hCH\}\approx 0.
\end{split}
\label{eqhCH}
\ee
The first two equations can be used to fix the Lagrange multipliers for the ``off-diagonal"
Gauss and diffeomorphism constraints\footnote{In the following equation,
the Poisson brackets with a slight abuse of notation denote the expressions appearing
under the integral in the part of the commutator non-vanishing on the constraint surface.
In other words, we consider the Poisson brackets of the non-smeared constraints dropping the distributional $\delta$-factor.}
\beq
G^a &=& \(\Nt\{\CH,\hCG_i\}+\Gt\{\hCH,\hCG_i\}\)\{\hCG_i,\hCD_a\}^{-1},
\nonumber\\
\tpsi^i &=& \Nt\(\{\CH,\hCG_j\}\{\hCG_j,\hCD_b\}^{-1}\{\hCD_b,\hCD_a\}+\{\CH,\hCD_a\}\)\{\hCD_a,\hCG_i\}^{-1}
\label{Lagrmult-found} \\
&&
+\Gt\(\{\hCH,\hCG_j\}\{\hCG_j,\hCD_b\}^{-1}\{\hCD_b,\hCD_a\}+\{\hCH,\hCD_a\}\)\{\hCD_a,\hCG_i\}^{-1}.
\nonumber\eeq
Plugging these expressions into the third equation, one observes that the terms proportional to $\Nt$ cancel and
the whole expression is proportional to $\Gt$. As a result, it leads to the following secondary constraint
\be
\begin{split}
\Psi=&\,\{\CH,\hCH\}+\{\CH,\hCG_i\}\{\hCG_i,\hCD_a\}^{-1}\{\hCD_a,\hCH\}
+\{\CH,\hCD_a\}\{\hCD_a,\hCG_i\}^{-1}\{\hCG_i,\hCH\}
\\
&\,
+\{\CH,\hCG_j\}\{\hCG_j,\hCD_b\}^{-1}\{\hCD_b,\hCD_a\}\{\hCD_a,\hCG_i\}^{-1}\{\hCG_i,\hCH\}.
\end{split}
\label{genPsi}
\ee
On the surface of this constraint, the last equation in \eqref{eqhCH} is satisfied automatically.
Thus, $\Psi$ is the only secondary constraint arising at this stage of the Dirac's procedure.

Having calculated in \eqref{hCD-hCH} all non-vanishing Poisson brackets entering the expression for $\Psi$,
one can proceed evaluating this constraint explicitly. This is done in appendix \ref{ap-secondcon}
where the following result is obtained:
\be
\Psi=-4\alpha\[\(e\(\tE^a_i-\tH^a_j\Et_b^j\tH^b_i\)+h\(\tH^a_i-\tE^a_j\Ht_b^j\tE^b_i\)\)\eta_a^i+\Delta^{-1} \Upsilon\],
\label{resPsi}
\ee
where
\be
\label{defvarie}
\begin{split}
& \Delta=(e-hT)e-(h-e\hT)h, \\
e=\det \tE^a_i, \qquad & h=\det \tH^a_i,
\qquad
T=\Ht_a^i\tE^a_i, \qquad \hT=\Et_a^i\tH^a_i,
\end{split}
\ee
and
\beq
\Upsilon &=&
\[e\(\tE^a_k\tE^b_k+\tH^a_k\tH^b_k\)+(h-e\hT)\tH^a_k\tE^b_k\]\epst_{bcd}\tE^c_l\tH^d_l
\epst_{agf}\eta_r^i\(\tH^g_i\(\tE^f_j\tE^r_j-\tH^f_j\tH^r_j\)+\tE^r_i\tE^g_j\tH^f_j\)
\nonumber\\
&&
-\[h\(\tE^a_k\tE^b_k+\tH^a_k\tH^b_k\)+(e-hT)\tH^a_k\tE^b_k\]\epst_{bcd}\tE^c_l\tH^d_l
\epst_{agf}\eta_r^i\(\tE^g_i\(\tE^f_j\tE^r_j-\tH^f_j\tH^r_j\)+\tH^r_i\tE^g_j\tH^f_j\)
\nonumber\\
&&
+\eps_{abr}\eps^{ijk}\eta^j_s\(\tE^r_i\tH^s_k-\tH^r_i\tE^s_k\)
\epst_{gcd}\(\tE^a_{k'}\tE^g_{k'}+\tH^a_{k'}\tH^g_{k'}\)\tE^c_l\tH^d_l\epst_{fpq}\tH^b_m\tE^f_m\tE^p_n\tH^q_n.
\label{Upsil}
\eeq
Although the result looks very complicated, the constraint features the simple properties of being linear in
the field $\eta$, and not containing any derivatives nor dependence on the connection $A$.
The expression \eqref{Upsil} can be further manipulated using various identities
and eliminating the epsilon tensors, but the result contains a number of monomials and is not particularly enlightening.
Finally, it is interesting to note that $\Upsilon$ is proportional to
the combination $\epst_{abc}\tE^b_i\tH^c_i$, analogous to the antisymmetric combination of the tetrads \eqref{eesym}
which must vanish for having agreement with the metric formulation. Therefore, on configurations satisfying
this condition, the secondary constraint crucially simplifies.

\subsection{Stability condition for the secondary constraint}
\label{subsec-stability}

The next step is to study the stability of the secondary constraint $\Psi$.
However, due to its complicated expression, it is difficult to do this by a direct computation.
Nevertheless, some important conclusions can be made if one realizes that $\Psi$ is given by
a ``partial Dirac bracket" of the two Hamiltonian constraints.
By partial Dirac bracket we mean here the Dirac bracket constructed using the minimal set of
non-commuting primary constraints. Since generically this is just a subset of the full set
of second class constraints of the theory, it is not yet the final Dirac bracket.
Nevertheless, it is a useful object since the final Dirac bracket is obtained from it
using the missing non-commuting constraints in the same way as it is constructed from Poisson bracket
using all second class constraints.

In our case, the minimal set of non-commuting primary constraints consists of $\hCG$ and $\hCD$.
Their Dirac matrix and its inverse are given by
\be\nonumber
\mathbb{D}_{ss'}=
\( \begin{array}{cc}
0 &  \{\hCG_i,\hCD_b\}
\\
\{\hCD_a,\hCG_j\} & \{\hCD_a,\hCD_b\}
\end{array}\),
\quad
\mathbb{D}^{ss'}=
\( \begin{array}{cc}
-\{\hCG_i,\hCD_a\}^{-1}\{\hCD_a,\hCD_b\}\{\hCD_b,\hCG_j\}^{-1} &  -\{\hCG_i,\hCD_b\}^{-1}
\\
-\{\hCD_a,\hCG_j\}^{-1} & 0
\end{array}\).
\ee
The partial Dirac bracket is the standard Dirac's formula
\be
\{A,B\}_D'=\{A,B\} -\{A,\Csc_s\}\mathbb{D}^{ss'}\{\Csc_{s'},B\},
\ee
where $\Csc_s=(\hCG_i,\hCD_a)$.
It is immediate to see that the expression \eqref{genPsi} for the secondary constraint is equivalent to
\be
\Psi=\{ \CH_1,\hCH\}_D',
\label{DirPsi}
\ee
where we smeared the Hamiltonian constraint $\CH$ with the trivial smearing function 1 just to remove
the $\delta$-function factor from the right-hand-side. Besides, it is useful to note
that the total Hamiltonian, after plugging in the expressions \eqref{Lagrmult-found} for the Lagrange multipliers
conjugate to the constraints $\Csc_s$, can be written as
\be
H_{\rm tot}=-\CG_{\tphi}-\CD_{\vN}-\CH'_N-\hCH'_G,
\label{Htot2}
\ee
where $\CH'$, $\hCH'$ are the two Hamiltonian constraints corrected by $\Csc_s$ such that
they weakly Poisson commute with all other primary constraints.
Due to this, the stability condition for $\Psi$ takes the following form
\be
\dot\Psi=\{\CH_N,\Psi\}_D'+\{\hCH_G,\Psi\}_D'\approx 0,
\label{stabPsi}
\ee
where we used the fact that $\Psi$ weakly commutes with $\CG_i$ and $\CD_a$ as well as
that under the partial Dirac bracket $\CH'$ and $\hCH'$ can be replaced by the original constraints.

The stability procedure ends if \Ref{stabPsi} can be interpreted as an equation fixing one of the Lagrange multipliers.
A necessary condition for this is that it should be algebraic in $\Nt$ and $\Gt$, i.e. it should not contain
their spatial derivatives. It turns out that this property follows just from the fact that $\Psi$ is well defined,
i.e. that $\{ \CH_N,\hCH_G\}_D'$ does not contain derivatives of the smearing functions.
Indeed, let us write the most general form of the commutator
\be
\{\CH_N,\{\CH_1,\hCH\}_D'\}_D'=\CX^a \p_a \Nt +\Nt \CY,
\label{genformcom}
\ee
where $\CX^a$ and $\CY$ are functions on the phase space.
Then, using Jacobi identity, one obtains
\be
\{\CH_N,\Psi\}_D'=\{\CH_1,\{\CH_N,\hCH\}_D'\}_D'+\{\hCH,\{\CH_1,\CH_N\}_D'\}_D'
\approx \Nt\(\CX^a \p_a 1 + \CY\)=\Nt \CY,
\ee
where at the second step we used \eqref{genformcom} in the first term and the fact that
any Dirac bracket with a second class constraint vanishes to remove the second contribution.
A similar computation can be done for the second term in \eqref{stabPsi} if one notice that \eqref{DirPsi}
can be equivalently rewritten as $\Psi=\{ \CH,\hCH_1\}_D'$.
Thus, we proved that the stability condition \eqref{stabPsi} for the secondary constraint
does not contain derivatives of $\Nt$ and $\Gt$,
and it is of the form
\be
\Nt \CY+\Gt \hCY=0.
\label{eqYY}
\ee

For generic configurations in phase space, the functions $\CY$, $\hCY$ are non-vanishing,
hence the stability condition \eqref{eqYY}
of the secondary constraint is a linear algebraic relation between the Lagrange multipliers $\Nt$ and $\Gt$.
As a result, the Dirac's procedure stops at this point, there are no tertiary constraints,
and the secondary constraint $\Psi$ is second class conjugate to one of the two Hamiltonian constraints.
This provides the complete description of the canonical structure of our model in generic case.

One may consider however the possibility of having configurations in phase space
for which the two factors in \eqref{eqYY} vanish simultaneously on the constraint surface,
\be
\CY=\{\CH_1,\Psi\}_D'\approx 0,
\qquad
\hCY=\{\hCH_1,\Psi\}_D'\approx 0.
\label{vanishDB}
\ee
If this is the case, $\Psi$ becomes weakly commuting with other constraints and thus
first class, further reducing the number of degrees of freedom.
Although for a generic phase space configuration this does not happen,
there still can exist some subsectors where \eqref{vanishDB} holds.
In fact, in appendix \ref{ap-pert} we show that such a possibility is indeed realized
and there are common solutions to the total set of constraints and the conditions \eqref{vanishDB}.
In particular, we study a perturbative expansion around $\tH^a=0$, evaluate explicitly the secondary constraint
in the quadratic approximation, and the stability condition to the linear order in $\tH^a$.
A common solution is shown to exist in the sector with $\tH^a=0$ and the connection $A_a$ restricted
to have a constant curvature set by the mass parameter $\alpha$.
However, the subsectors thus found appear to be not stable under the evolution,
which makes impossible to interpret the conditions \eqref{vanishDB} as consistent additional constraints.
It is likely that they do not have a physical significance and arise due to our incomplete understanding
of the decoupling of the diagonal and off-diagonal degrees of freedom at the non-linear level.

\subsection{Summary}
\label{subsec-sum}

Let us summarize what we have found so far. Generically the theory possesses 15 constraints. Among
them there are 7 first class constraints: $\CG_i$, $\CD_a$ and $\CH'_N+\hCH'_G$, with
the Lagrange multipliers $\Nt$ and $\Gt$ related by the condition \eqref{eqYY}.
These constraints generate the local symmetries of the chiral Plebanski formulation of general relativity:
$SU(2)$ gauge transformations, spatial and time diffeomorphisms.
The remaining 8 constraints are of second class: $\hCG_i$, $\hCD_a$, $\Psi$ and either $\hCH$ or $\CH$
depending which of them does not commute with $\Psi$.\footnote{If they both do not commute with the secondary constraint,
any of them can be chosen as second class. This is because they can be related to each other
by adding a first class constraint, which represents a general ambiguity in the Dirac's approach.}
Altogether these constraints fix $2\times 7+8=22$ variables in the initial $4\times 9=36$ dimensional phase space.
As a result, the theory turns out to describe 7 degrees of freedom, and it is free from the Boulware-Deser ghost.

Expanding around a bi-flat background, the degrees of freedom can be identified with a massless
and a massive spin-2 fields. Specifically, the massless graviton is encoded
in the ``diagonal" $(\tE,A)$ sector supplemented by
7 first class constraints, whereas the 5 degrees of freedom of the massive one
are carried by the traceless symmetric part of the ``off-diagonal" $(\tH,\eta)$ sector,
with the antisymmetric and trace parts being fixed by the second class constraints
(see appendix \ref{ap-pert}).
As anticipated, the identification of the physical fields at the non-linear level is more complicated.
An explicit calculation in appendix C shows that the physical spatial metric identified by
the Poisson bracket between two Hamiltonians is quite a non-trivial function
lacking a simple expression in terms of our basic variables.

It is interesting to compare the resulting canonical structure of this model with
the one of bi-metric gravity in the metric formalism. In the latter case, one starts with two uncoupled copies of gravity, with the group of gauge symmetries
being two copies of the group of diffeomorphisms ${\it Diff}$, and as a result 2+2 propagating
degrees of freedom.  After the two copies of gravity are coupled, the group of gauge symmetries
is broken down to the diagonal group of diffeomoprhisms. This removes 4 of the gauge symmetries
of the original uncoupled theory, and thus adds 4 propagating degrees of freedom.
If, however, the interaction term is tuned to be of a special form such that it is linear in the two lapse
functions, in addition one generates a couple of second class constraints (one primary and one secondary)
removing one degree of freedom corresponding to the Boulware-Deser ghost and leaving us with 7 degrees of freedom.

A similar mechanism is at play in our model as well, but with some important differences.
Like in the metric formulation, the group of gauge symmetries of
the two uncoupled copies of Plebanski theory is also broken down to the diagonal subgroup
by the interaction term. However, as \eqref{inttermsplit} shows, the interaction term is
linear in both, lapse and shift functions. Thus, while
the off-diagonal group of gauge transformations is broken, there are still primary
constraints corresponding to these transformations. The primary constraints for off-diagonal
spatial diffeomorphisms together with the primary constraints for off-diagonal ${\rm SO}(3)$
rotations then turn out to form a second-class pair. Thus, even though by breaking
the spatial diffeomorphisms and ${\rm SO}(3)$ rotations we have removed 3+3 gauge symmetries,
we have only added 3 degrees of freedom. To say it differently, the interaction term just converted
the corresponding constraints from first class to second class.
As a result, they now remove a ``half" of the configurational degree of freedom each,
as compared to removing one when they are first class.
On the other hand, the removal of the Boulware-Deser ghost proceeds in the way analogous to the metric
formulation and is based on the existence of the additional second class pair.
While very similar picture was outlined already in  \cite{Hinterbichler:2012cn},
in this paper we have given its explicit realization for the first time.

As we have already remarked above, our model is equivalent to the metric formulation
only on the configurations satisfying the symmetry condition \eqref{eesym}.
The natural question is then if one can impose this condition by hand,
i.e. with the help of some Lagrange multiplier terms added to the action.
It is not hard to convince oneself that this is not possible,
because this condition cannot be rewritten in terms of the $B$ fields.
However, as we already discussed in section \ref{sec-massive}, the condition \eqref{eesym}
also appears if one requires the reality of the action \eqref{ourtheory} once the standard
reality conditions on the $B$ fields are imposed. It hints that \eqref{eesym} should arise
in this case as a dynamical constraint. This expectation indeed turns out to be true since
the spatial part of \eqref{eesym} coincides then with the real part of $\hCD$.
However, the full canonical analysis of the action \eqref{ourtheory} supplemented by
the standard reality conditions leads
to a disappointing conclusion that such a system describes just two massless gravitons
coupled by a gauge fixing term. This happens because there are too many constraints and
the massive graviton modes disappear. The counting of degrees of freedom in this case works as follows.
The diagonal sector clearly describes two degrees of freedom, so we discuss only the off-diagonal part.
It is given by $2\times 2\times 9=36$ real dimensional phase space of complex fields $(H,\eta)$
supplemented by 18 reality conditions and by $2\times 8=16$ constraints $\hCG_i$, $\hCD_a$, $\hCH$ and $\Psi$.
However, it can be checked that 5 of these constraints, namely $\Re\hCG_i$, $\Im\hCH$ and $\Re\Psi$,
coincide with the reality conditions on the surface of other constraints. Moreover,
one should take into account that in the real theory not all of the above constraints
are second class: $\Im \hCD$ is in fact first class. As a result, after imposing all constraints,
we get $36-18-8-2\times 3=4$ dimensional phase space corresponding to the degrees of
freedom of a second massless graviton. Thus, imposing the standard reality conditions,
and generating the symmetry condition \eqref{eesym} as a dynamical constraint,
does not provide a way to relate our model to the metric formulation of massive gravity.
If such a relation exists, it should rely on a more sophisticated choice of reality conditions.

Finally, we have also observed that there are sectors in phase space,
characterized by vanishing of the two (partial Dirac) brackets \eqref{vanishDB},
where this classification of constraints may fail, in particular, leading to fewer degrees of freedom.
In fact, the existence of special sectors in phase space with a drastically different  canonical structure
is not an unusual situation. It happens, for example, in the first order tetrad formulation
of general relativity, and as well in the Plebanski formulation, where one can allow degenerate tetrads
with the vanishing determinant of the metric.
For instance, the degenerate sector of the non-chiral Plebanski formulation corresponds to a topological
theory with no local degrees of freedom \cite{Alexandrov:2012pj}.
In principle, something similar might happen in our case as well,\footnote{Similar conclusions have
been made in \cite{Kluson:2012gz} for a model of non-linear massive gravity in St\"uckelberg formalism.}
but the geometric interpretation of such subsectors is far from clear.
In any case,  they can lead only to a reduction of degrees of freedom,
and therefore our analysis ensures that the scalar ghost has been removed throughout all the phase space.

\subsection{Modified Plebanski theory}
\label{subsec-mod}
A striking property of the Plebanski formulation of general relativity is that it allows
an extension to an infinite class of actions all propagating only 2 degrees of freedom.
This should be compared with traditional modifications of general relativity like higher-order
$f(R)$ theories, which typically have additional degrees of freedom. Such special modification
is achieved by turning the cosmological constant in \Ref{Plebact} into an arbitrary functional $\l(\Psi)$
of the Lagrange multiplier $\Psi$. The resulting action describes a modified theory of gravity
propagating two degrees of freedom only \cite{Krasnov:2007cq}.
It can be argued to be related to an infinite sum of higher order curvature terms, organized
as to not propagate extra degrees of freedom \cite{Krasnov:2009ik}.
The specific form of the infinite summation depends on the specific choice of $\l(\Psi)$.

Relying on these results, one can suggest a similar generalization of our model \eqref{ourtheory}
where the two cosmological constants are replaced by arbitrary functionals $\l^\pm(\Psi^\pm)$.
The argument implying absence of extra degrees of freedom goes through for both sectors
of the coupled system, and the interaction does not appear to spoil it.
Moreover, the interaction term is still linear in both lapses and shifts, which indicates the presence
of the constraints responsible for the absence of the Boulware-Deser ghost.
Therefore, even without performing a detailed canonical analysis, we argue that there
is in fact an infinite class of ghost-free bi-metric gravities, characterized by the functionals $\l^\pm(\Psi^\pm)$.
This might be particularly useful concerning phenomenological applications of the theory.
We leave the study of these issues for future research.

\section{Conclusions}
\label{sec-concl}

In this paper we have presented a chiral model for ghost-free massive (bi-)gravity.
The theory we considered is given by two copies of the Plebanski action for general relativity,
plus an interaction term. On shell, the interaction corresponds to
the sum of two terms \Ref{BB}, one of which coincides with the symmetric interaction between tetrads considered
in \cite{Hinterbichler:2012cn} (second item of \Ref{list}) and the other features
an alternative gauge-invariant contraction of the tetrads.
We performed a complete canonical analysis and proved that the theory only propagates seven degrees of freedom,
hence effectively removing the Boulware-Deser ghost usually plaguing models of massive gravity.
Thus, our results give an independent and explicit proof of the absence of the ghost
for a special type of interaction terms \cite{deRham:2010ik,deRham:2010kj,Hassan:2011tf,Hassan:2011zd},
and furthermore extend it for the first time to a first order action of gravity,
with independent tetrad and connection variables.

It is instructive to compare and contrast the formulation of bi-metric gravity used here with the other,
more standard ones such as metric and tetrad formulations.
The former is the most economic one in terms of fields, as any gauge formulation introduces
unphysical redundancies, and both the tetrad and chiral Plebanski-type formulations have more of these.
However, the interaction terms needed to remove the ghost are non-polynomial in the metric framework,
which makes computations increasingly complex.
The tetrad formulation converts these complicated interactions into simple polynomials,
but at the same time it introduces the additional gauge symmetry of local frame rotations
without really simplifying the Hamiltonian analysis.
What we have considered in this paper is a chiral model, which lies in between the pure metric and the tetrad formulations
since it brings only 3 local gauge rotations as compared to 6 in the tetrad formalism.
The reduction of the gauge group tremendously simplifies the canonical analysis,
a fact that can be exploited to perform a number of explicit calculations.

Our formalism does not allow us to incorporate the first and third terms
in \Ref{list}, hence it may look more restricting, allowing de facto only
the second term in the list. However, as we discussed, the use of
2-forms as fundamental variables also opens the door to the possible existence of
an \emph{infinite} class of ghost-free theories, by turning the cosmological constants $\l^\pm$ into arbitrary
functionals of the Lagrange multipliers $\Psi^{\pm}$. These in turn introduce new interactions
for the metrics, and the procedure can be argued to not introduce extra degrees of freedom.
The existence of a more general class of ghost-free interaction is interesting, and deserves to be
further studied.\footnote{An infinite number of ghost-free theories can also be obtained if one considers
Lorentz-breaking potentials \cite{Comelli:2012vz} (see also \cite{Dubovsky:2004sg}
for more on Lorentz-breaking massive gravity).
Here instead it would be an infinite class of Lorentz-invariant potentials.}

The chiral formulation has a drawback in that the analysis of bi-metric theory
holds, strictly speaking, only for metrics with Euclidean signature. The calculations extend immediately
to the physical case of Lorentzian signature, however in this case one works from the start
with complex fields, to be supplemented by appropriate reality conditions.
The simplest possibility is to borrow the usual reality conditions for Lorentzian Plebanski
gravity and apply them independently to both sets of fields. This \emph{almost} works:
it induces real and Lorentzian metrics, and the reality conditions are preserved
by time evolution. However, the total action is not made real by these conditions,
unlike in the simple gravity case. One imaginary term survives
and spoils the theory, and this is precisely the parity-odd extra term in \Ref{BB}.
Interestingly, the term that makes the standard reality conditions fail is
the same that characterizes the difference between bi-gravity in the metric and tetrad formulations.
Hence, finding more non-trivial reality conditions could also be related to finding
a reformulation of the Plebanski version  totally equivalent to the metric one.

A way to avoid the problem of reality conditions would be to start with
the non-chiral Plebanski theory based on the full Lorentz group \cite{Capovilla:1991qb,MikeLR,DePietri:1998mb},
which is at the basis of the spin foam quantization program to general relativity
\cite{Perez:2003vx,Alexandrov:2011ab}.
However, its canonical analysis is of the same difficulty as the one in the tetrad formulation,
see \cite{Buffenoir:2004vx,Alexandrov:2006wt}. Furthermore, modifying the non-chiral Plebanski theory
in a way similar to the one considered in section \ref{subsec-mod}, in contrast to the chiral case,
one obtains a theory with generically 8 degrees of freedom \cite{Alexandrov:2008fs},
that can be interpreted as those of a generic massive gravity
including the scalar ghost \cite{Speziale:2010cf}. This suggests an interesting program of realizing
the ghost-free massive gravity as a particular subclass of modified (non-chiral) Plebanski theories.
An investigation along these lines has appeared in \cite{Beke:2011mu}.

Another interesting open question concerns the existence of stable subsectors with fewer
degrees of freedom. Apart from the trivial cases of $\alpha=0,\infty$ (corresponding
respectively to two decoupled gravity theories and a single one), our analysis allows
us to explicitly characterize such subsectors by analyzing the stability condition for the secondary constraint.
They appear unstable at first sight, but a more complete study of their
dynamical properties and geometric interpretation is needed to arrive at a conclusive result.

Finally, it would be interesting to see whether quantization techniques,
such as loop quantization and spin foam approach, developed to deal with gauge formulations of gravity,
can be extended to the models describing massive (bi-)gravity.

\section*{Acknowledgements}
This work was supported by contract ANR-09-BLAN-0041, an ERC Starting Grant 277570-DIGT,
as well as partially by the Alexander von Humboldt foundation, Germany.

\bigskip

\section*{Appendix}
\appendix

\section{Evaluation of the secondary constraint}
\label{ap-secondcon}

In this appendix we evaluate the secondary constraint whose general form is given in \eqref{genPsi}.
The first term in this expression is just the Poisson bracket between the diagonal and off-diagonal
Hamiltonian constraints computed in \eqref{hCD-hCH}. It can be further simplified as
\be
\begin{split}
\{\CH,\hCH\}=&\,
-4\alpha\[e\(\tE^a_i-\tH^a_j\Et_b^j\tH^b_i\)+h\(\tH^a_i-\tE^a_j\Ht_b^j\tE^b_i\)\]\eta_a^i,
\end{split}
\ee
where $e=\det \tE^a_i$, $h=\det \tH^a_i$,
and thus reproduces the first term in \eqref{resPsi}.
The second term includes contributions of the remaining three terms in \Ref{genPsi}.
To evaluate them, the first step is to invert the matrix
\be
\Delta_a^i:=\f1{4\a}\{\hCD_a,\hCG_i\} = \hf\,\eps^{ijk} \epst_{abc} \(\tE^{b}_j\tE^{c}_k - \tH^{b}_j\tH^{c}_k\)= e \Et_a^i-h\Ht_a^i.
\ee
Its inverse can be computed as follows
\be
\begin{split}
\(\Delta^{-1}\)^a_i=&\,
\frac{1}{2\Delta}\,\eps^{ijk}\epst_{abc}\Delta_b^j\Delta_c^k
=\Delta^{-1}\[e\tE^a_i+h\tH^a_i - \hf \(\tE^a_i\tE^b_j-\tE^b_i\tE^a_j\)\eps^{jkl}\eps_{bcd}\tH^c_k\tH^d_l\],
\end{split}
\ee
where
\be
\Delta=\det \Delta_a^i=(e-hT)e-(h-e\hT)h,
\label{defDelta}
\ee
and $T=\Ht_a^i\tE^a_i$, $\hT=\Et_a^i\tH^a_i$.
Using this result, one further computes
\be
\begin{split}
\{\CH,\hCG_i\}\{\hCG_i,\hCD_a\}^{-1} =&\, -\Delta^{-1}\[e\(\tE^a_i\tE^b_i+\tH^a_i\tH^b_i\)
+(h-e\hT)\tH^a_i\tE^b_i\]\epst_{bcd}\tE^c_j\tH^d_j,
\\
\{\hCD_a,\hCG_i\}^{-1}\{\hCG_i,\hCH\} =&\, -\Delta^{-1}\[h\(\tE^a_i\tE^b_i+\tH^a_i\tH^b_i\)
+(e-hT)\tH^a_i\tE^b_i\]\epst_{bcd}\tE^c_j\tH^d_j,
\end{split}
\label{exp-comb}
\ee
which allows to get the last term in \Ref{genPsi} in the following form
\be
\begin{split}
& \{\CH,\hCG_j\}\{\hCG_j,\hCD_b\}^{-1}\{\hCD_b,\hCD_a\}\{\hCD_a,\hCG_i\}^{-1}\{\hCG_i,\hCH\}
\\
& \qquad
=4\alpha\Delta^{-1}\epst_{abr}\(\tE^r_i\p_s\tE^s_i-\tH^r_i\p_s\tH^s_i\)
\epst_{gcd}\(\tE^a_j\tE^g_j+\tH^a_j\tH^g_j\)\tE^c_k\tH^d_k\epst_{fpq}\tH^b_l\tE^f_l\tE^p_m\tH^q_m.
\end{split}
\label{brack1}
\ee
This contribution can be combined with $A$-dependent terms from
\be
\begin{split}
& \{\CH,\hCG_i\}\{\hCG_i,\hCD_a\}^{-1}\{\hCD_a,\hCH\}+\{\CH,\hCD_a\}\{\hCD_a,\hCG_i\}^{-1}\{\hCG_i,\hCH\}
\\
&\qquad
= -4\alpha\Delta^{-1}\[\(e\(\tE^a_i\tE^b_i+\tH^a_i\tH^b_i\)+(h-e\hT)\tH^a_i\tE^b_i\)A_a^j\tH^r_j
\right.
\\
&\qquad\quad
\left. -\(h\(\tE^a_i\tE^b_i+\tH^a_i\tH^b_i\)+(e-hT)\tH^a_i\tE^b_i\)A_a^j\tE^r_j\]
\epst_{bcd}\tE^c_k\tH^d_k\epst_{rgf}\tE^g_l\tH^f_l
\\
&\qquad\qquad
+{\mbox{terms linear in }\eta}.
\end{split}
\label{brack2}
\ee
One can check that these terms complete the derivatives in \eqref{brack1} to covariant ones, which in turn
can be replaced by the two Gauss constraints provided one subtracts the missing terms linear in $\eta$.
The contributions proportional to the Gauss constraints can be dropped, whereas
the terms linear in $\eta$ are combined with similar ones from \eqref{brack2} producing the last term in \eqref{resPsi}
explicitly given in \eqref{Upsil}.

\section{Perturbative analysis around $\tilde H^a=0$}
\label{ap-pert}

In this appendix we analyze in detail the secondary constraint and its stability condition
in a first few orders of a perturbative expansion around $\tH^a=0$. It should be emphasized that
this analysis is more general than a linearization around a flat background.
In particular, we will show that, in contrast to the later
(under the assumption of non-degeneracy of the physical metric),
it allows for solutions of the conditions \eqref{vanishDB} for which the stability condition of the secondary constraint
does not seem to impose a restriction on the Lagrange multipliers.

First, let us compute the secondary constraint $\Psi$ up to quadratic terms in $\tH^a$.
Taking into account that
\be
\Delta =e^2 +O(\tH^2),
\ee
from the explicit expression \eqref{resPsi} together with \eqref{Upsil} one can show that
the constraint can be written as
\be
\begin{split}
\Psi \approx&\,
-4\alpha e\[ \(1+O(\tH^2)\) \tE^a_i\eta_a^i +\eta_a^i\tH^a_i\Et_b^j\tH^b_j-\eta_a^i\tH^b_i\Et_b^j\tH^a_j
-\eta_a^i\tH^c_i\tE^a_j\tH^b_j (\Et\Et)_{bc} \] +O(\tH^3),
\end{split}
\ee
where the constraint $\hCG_i$ has been extensively used.
As a result, dividing by the prefactor of the first term, one concludes that
in our approximation it is sufficient to consider
\be
\hat\Psi=\tE^a_i\eta_a^i+ \hat\Psi^{(2)},
\qquad
 \hat\Psi^{(2)}=\eta_a^i\tH^a_i\Et_b^j\tH^b_j-\eta_a^i\tH^b_i\Et_b^j\tH^a_j
-\eta_a^i\tH^c_i\tE^a_j\tH^b_j (\Et\Et)_{bc}.
\label{Psitwo}
\ee
In particular, we observe that at the leading order the secondary constraint ensures
the vanishing of the trace part of the $\eta$ field. Similarly, using \eqref{constraints},
one can interpret the ``off-diagonal" Hamiltonian constraint $\hCH$ as a restriction
on the trace part of $\tH$ (since the last term gives $\alpha\Et_a^i\tH^a_i$),
whereas $\hCG_i$ and $\hCD_a$ can be viewed as restrictions on the antisymmetric parts of $\eta$ and $\tH$,
respectively.
This observation is already sufficient to conclude that in the $(\tH,\eta)$ sector we can have at most 5
propagating degrees of freedom, contained in the symmetric traceless parts of the fields, which
are expected to correspond to the degrees of freedom of a massive graviton.
The ghost scalar carried by the trace part is removed by the constraints $\hCH$ and $\Psi$.

Next, let us analyze the stability condition of the secondary constraint which requires
the vanishing of the Poisson bracket $\{\hat\Psi,H_{\rm tot} \}$. We will compute it
keeping only terms which are at most linear in $\tH^a$.
To this end, using \eqref{Lagrmult-found} and \eqref{exp-comb}, on the constraint surface we obtain
\be
\begin{split}
G^a =&\, -e^{-1}\Nt \tE^a_i\tE^b_i \epst_{bcd}\tE^c_j\tH^d_j+O(\tH^2),
\\
\tpsi_i \approx &\, \Nt\( D_a \tH^a_i -e^{-1} \epst_{abc}A_d^i\tE^d_j\tE^a_j\tE^b_k\tH^c_k\)
+e^{-1}\Gt\epst_{abc}\tE^a_i\tE^b_j\eta_d^k\(\tH^c_k\tE^d_j-\tH^c_j\tE^d_k\)+O(\tH^2).
\end{split}
\ee
Given the above results, a direct computation gives
\beq
\left\{\hat\Psi,H_{\rm tot}\right\}&\approx&
\left\{ \hCG_{\tpsi} +\hCD_{\vec G}+\CH_N+\hCH_G,\tE^a_i\eta_a^i\right\}
+\frac{\delta (\CH_N+\hCH_G)}{\delta\eta_a^i}\,\frac{\p \hat\Psi^{(2)}}{\p\tH^a_i}
+O(\tH^2)
\nonumber\\
&\approx&
\Nt \[ 2\eps^{ijk}\tE^a_i\tH^b_j\eps^{klm}\eta_a^l\eta_b^m
+3\alpha\eps^{ijk}\epst_{abc}\tE^a_i\tE^b_j\tH^c_k
+\eps^{ijk}\tE^a_i\tE^b_j\eps^{klm}\eta_b^m \frac{\p \hat\Psi^{(2)}}{\p\tH^a_l}\]
\nonumber
\\
&&
+\[-\Gt\eps^{ijk}\tE^a_i\tE^b_j \(F_{ab}^k-\alpha\epst_{abc}\tE^c_k\)
+D_b\(\Gt\eps^{ijk}\tE^a_i\tE^b_j \)\frac{\p \hat\Psi^{(2)}}{\p\tH^a_k}
\right.
\nonumber \\
&& \Biggl.
+\p_a\(\Gt\eps^{ijk}\tE^a_i\(\tE^b_j\Et_c^k\tH^c_l\eta_b^l-\tH^b_j\eta_b^k\)\)
\Biggr]+O(\tH^2).
\label{stab-Psi}
\eeq
Using \eqref{Psitwo} and the constraint $\CH$ in the second term,
on the constraint surface (including $\hat\Psi$) this can be further simplified as
\be
\begin{split}
\left\{ \hat\Psi, H_{\rm tot}\right\}\approx&\,
\Nt  \(\(6\alpha e-\tE^a_i\eta_b^i\tE^b_j\eta_a^j \)\Et_c^k\tH^c_k+2\tE^a_i\eta_b^i\tH^b_j\eta_a^j\)
\\
&\,
+\Gt \(6\alpha e-\tE^a_i\eta_b^i\tE^b_j\eta_a^j  +W_H \)+O(\tH^2),
\end{split}
\label{firstcoef}
\ee
where
\be
\begin{split}
W_H =&\, 2\eps^{ijk}\tE^a_i\tH^b_j\(D_a\eta_b^k-D_b\eta_a^k\)
-\eps^{ijk}\tE^a_i\(\eta_d^l\tH^d_l \Et_b^k D_a\tE^b_j-\tE^b_j\Et_d^l\tH^d_l D_a\eta_b^k
\right.
\\
&\, \left.
+\Et_d^j\tH^d_l\eta_b^l D_a\tE^b_k  +2\tE^b_j\tE^d_k\eta_d^l\tH^c_l\Et_c^m D_a\Et_b^m
+\tE^b_j D_a(\eta_d^k\Et_b^l\tH^d_l)+D_a(\eta_d^k\tE^d_l\tH^c_l\Et_c^j)
\).
\end{split}
\ee
In particular, all terms with derivatives of the Lagrange multipliers cancel in agreement
with the general conclusion of section \ref{subsec-stability}.

Both coefficients, in front of $\Nt$ and $\Gt$, appear to be non-trivial functions on the phase space.
If at least one of them is non-vanishing, the equation $\{\hat\Psi,H_{\rm tot}\}=0$ can be considered
as an equation for the corresponding Lagrange multiplier. Then the Dirac's procedure is completed
and $\Psi$ together with one of the Hamiltonian constraints is of second class.

Let us now analyze the situation where both coefficients in \eqref{firstcoef} simultaneously vanish.
One might expect that in this case the constraint structure changes drastically.
In particular, there is a chance that both Hamiltonian constraints and $\Psi$ become first class.
In such situation, although the ghost degree of freedom is still removed, it would happen
due to an additional gauge symmetry, rather than by means of constraints which are of second class.
The perturbative expansion presented in this appendix provides a framework to explicitly address these issues.

First, we need to understand whether it is really possible to have both coefficients vanishing.
Explicitly this requires
\be
\begin{split}
\CY:= &\, \(6\alpha e-\tE^a_i\eta_b^i\tE^b_j\eta_a^j \) \Et_c^k\tH^c_k+2\tE^a_i\eta_b^i\tH^b_j\eta_a^j=0,
\\
\hCY:=&\,6\alpha e- \tE^a_i\eta_b^i\tE^b_j\eta_a^j +W_H=0.
\end{split}
\label{defYY}
\ee
To start with, we consider a linearization around a flat background which can be described by
$\tE^a_i=\delta^a_i$ and
$A_a^i=0$.
Using the flatness of the connection, from the Hamiltonian constraint $\CH$ \eqref{constraints} one finds that
\be
\tE^a_i\eta_b^i\tE^b_j\eta_a^j=O(\tH).
\ee
Combined with $\hCY$ in \eqref{defYY}, this leads to $6\alpha=O(H)$ which is inconsistent with our assumption of
$\tH^a$ being infinitesimally small. Thus, near a flat background $\hCY\ne 0$ so that $\hCH$ forms a second class system with
the secondary constraint $\Psi$, and the model describes one massless and one massive graviton,
consistently with previous findings \cite{Hinterbichler:2012cn}.

The above consideration suggests that one should look for a common solution to \eqref{defYY} and all other constraints
in the sector with a constant curvature set by the mass parameter $\alpha$, rather than the flat one.
Let us do this setting for simplicity $\tH^a=0$ which automatically ensures $\CY=0$.
In the sector with vanishing $\tH^a$ the constraints become
\be
\begin{split}
&
\hat\Psi= \tE^a_i\eta_a^i,
\qquad
\CG_i = D_a \tE^{a}_i,
\qquad
\hCG_i = \eps_{ijk} \eta^j_a \tE^{ak} ,
\\
&
\CC_a= \tE^{b}_i (F_{ab}^i + \eps^{ijk} \eta_a^j \eta_b^k),
\qquad
\hCC_a=  \tE^{b}_i (D_a \eta_b^i - D_b \eta_a^i),
\\
&
\CH=  \frac{1}{2}\,\eps^{ijk} \tE^{a}_i\tE^{b}_j(F_{ab}^k + \eps^{klm} \eta_a^l \eta^m_b ),
\qquad
\hCH= \eps^{ijk} \tE^a_i\tE^{b}_j D_a \eta_b^k.
\end{split}
\label{constraintsH0}
\ee
Introducing $\eta^{ij}=\eta^i_a\tE^a_j$, and requiring that it is symmetric and traceless, solves $\hCG_i$ and $\Psi$
so that we remain with the following equations
\be
\begin{split}
&
\tilde\eps^{abc}\Et_a^i F_{bc}^i=6\alpha,
\qquad
D_a \tE^{a}_i=0,
\\
&
\tE^{b}_i F_{ab}^i=0,
\qquad
\tE^{b}_i \eta^{ij}(D_a \Et_b^j - D_b \Et_a^j)-\tE^{b}_i\Et_a^jD_b\eta^{ij}=0,
\\
&
\eta^{ij}\eta_{ij}=6\alpha,
\qquad
\eta^{ij}\tilde\eps^{abc} \Et_c^i  D_a \Et_b^j=0.
\end{split}
\label{eqsH0}
\ee
The first two of these equations require the connection to be consistent with the triad and
to have a constant curvature, whereas the fifth equation fixes the ``norm" of the field $\eta^{ij}$.
Then we remain with 5 differential equations on 9 components of $\tE^a_i$.
It is natural to expect that they can be simultaneously solved which shows that the space
of common solutions to the conditions \eqref{defYY} (or more generally to \eqref{vanishDB})
and the constraints is not empty.

The next step is to study the stability of the additional constraints \eqref{defYY}.
Our approximation allows to compute the commutators in the zeroth order in $\tH^a$.
Explicitly, one finds
\be
\begin{split}
\{\CY,H_{\rm tot}\}\approx &\,
\{\CH_N+\hCH_G,\CY\}+O(\tH)
\approx
-2\Nt {\eta^i}_j{\eta^j}_k{\eta^k}_i+2\Gt\eps^{ijk}\tE^a_i\tE^b_j\eta_{kl}\eta^{lm} D_a\Et_b^m
+O(\tH),
\\
\{\hCY,H_{\rm tot}\}\approx &\,
\{\CH_N+\hCH_G,\hCY\}+O(\tH)
\approx
2\Nt\eps^{ijk}\tE^a_i\tE^b_j\eta_{kl}\eta^{lm} D_a\Et_b^m
\\
&\, +\Gt\[2\eps^{ijk}\tE^a_i\eta_j^l \tE^b_l F_{ab}^k
-\left\{\int \de^3y \, \eps^{ijk}D_a(\tE^a_i\tE^b_j)\eta_b^k ,W_H\right\}\]
+O(\tH),
\end{split}
\ee
All terms here appear to be non-vanishing. As a result, the stability of $\CY$ and $\hCY$
cannot be achieved unless both Lagrange multipliers $\Nt$ and $\Gt$ simultaneously vanish.
However, we are not interested in such situation as it corresponds to degenerate metrics.
Therefore, the sector defined by \eqref{defYY} does not seem to possess an interesting dynamics
and probably does not have any physical significance.\footnote{An alternative possibility to treat
the conditions \eqref{defYY} would be to insert them into the original Lagrangian.
Then they appear in the total Hamiltonian with appropriate Lagrange multipliers and
their stability should be studied at the same step as the stability of the primary constraints.
In this scenario, the stability of $\CY$ and $\hCY$ is ensured by fixing the Lagrange multipliers and does not require
the vanishing of $\Nt$ and $\Gt$. However, due to
$$
\{ \CY(x) ,\hCY(y)\}\approx 4{\eta^i}_j{\eta^j}_k{\eta^k}_i \,\tilde\delta^3(x-y)+O(\tH),
\label{comYY}
$$
the constraints $\CY$ and $\hCY$ appear to be second class and affect the expression of the secondary constraint.
As a result, this scenario leads to a constraint surface which does not intersect with the one corresponding
to the physically interesting case of massive gravity.}

\section{Commutator of two Hamiltonian constraints}
\label{ap-twoH}

Let us consider the commutator of the two first class constraints playing the role
of the Hamiltonian constraint. Such constraint can be obtained by collecting all terms in the total Hamiltonian \eqref{Htot1}
which are proportional to the ``diagonal" lapse function $\Nt$ after plugging in all solutions for
the Lagrange multipliers fixed by the stability conditions. Thus, it is given by
\be
\CH_N^{\rm f.c.}=\CH_N+\hCH_{G(N)}+\hCD_{\vec G(N)}+\hCG_{\tilde\psi(N)},
\ee
where $\Gt(N), G^a(N), \tilde \psi^i(N)$ are those solutions which can be found from
\eqref{Lagrmult-found} and \eqref{eqYY}.
Our aim here is to get the function analogous to $N^a(N,M)$ in \eqref{notforalgebra}
to read off the physical metric determined by the diffeomorphism algebra.
Therefore, we are not interested in the full commutator, but only in the terms
proportional to the diffeomorphism constraint $\CD$.
Using the constraint algebra presented in \eqref{fullalgebra}, one finds
\be
\{\CH_{N_1}^{\rm f.c.},\CH_{N_2}^{\rm f.c.}\}=\CV^a(N_1,N_2) \CD_a +\cdots,
\ee
where
\be
\begin{split}
\CV^a(N_1,N_2)=&\, V^a(N_1,N_2)+V^a(G(N_1),G(N_2))-U^a(N_1,G(N_2))-U^a(G(N_1),N_2)
\\
&\,
+L^a(\vec G(N_1),\vec G(N_2)).
\end{split}
\ee
Let us take
\be
G^a(N)=\Nt f^a+\Gt(N) g^a, \qquad
\Gt(N)=\CR \Nt,
\ee
where our results imply that
\be
f^a=\{\CH,\hCG_i\}\{\hCG_i,\hCD_a\}^{-1},
\qquad
g^a=\{\hCH,\hCG_i\}\{\hCG_i,\hCD_a\}^{-1},
\qquad
\CR=-\CY/\hCY.
\ee
Then one easily calculates that
\be
\CV^a(N_1,N_2)=\CK^{ab}\(\Nt_1\p_b\Nt_2-\Nt_2\p_b\Nt_1\),
\ee
where
\be
\begin{split}
\CK^{ab}=&\,
(1+\CR^2)\((\tE\tE)^{ab}+(\tH\tH)^{ab}\)-2\CR\(\tE^a_i\tH^b_i+\tE^b_i\tH^a_i\)+(f^a+\CR g^a)(f^b+\CR g^b)
\\
=&\, (\tE^a_i-\CR\tH^a_i)(\tE^b_i-\CR\tH^b_i)+(\tH^a_i-\CR\tE^a_i)(\tH^b_i-\CR\tE^b_i)
+(f^a+\CR g^a)(f^b+\CR g^b).
\end{split}
\label{resKab}
\ee
The function $\CK^{ab}$ is expected to encode the spatial part of the physical metric of the coupled theory as
$\CK^{ab}=g g^{ab}$. The result \eqref{resKab} clearly shows that the physical metric is a very complicated function
being expressed in terms of ``diagonal" and "off-diagonal" variables.

Using results from appendix \ref{ap-pert}, it is possible to get a more explicit
representation for $\CK^{ab}$ in the quadratic approximation in $\tH^a$.
One finds
\be
\begin{split}
\CK^{ab}\approx &\,
(\tE\tE)^{ab}\[1+\(\Et_c^i\tH^c_i+\frac{2\tE^c_i\eta_d^i\tH^d_j\eta_c^j}{6\alpha e-\tE^g_k\eta_f^k\tE^f_l\eta_g^l}\)^2\]
+(\tH\tH)^{ab}
\\
&\,
+2\(\tE^a_m\tH^b_m+\tE^b_m\tH^a_m\)\(\Et_c^i\tH^c_i+\frac{2\tE^c_i\eta_d^i\tH^d_j\eta_c^j}{6\alpha e-\tE^g_k\eta_f^k\tE^f_l\eta_g^l}\)
\\
&\,
+\eps^{ijk}\tE^a_i\tH^c_j\Et_c^k\eps^{lmn}\tE^b_l\tH^d_m\Et_d^n +O(\tH^3).
\end{split}
\ee
Expanding around a bi-flat background
\be
\tE^a_i=\delta^a_i+f^a_i,
\qquad
f\sim A\sim \tH\sim \eta\sim o(1),
\ee
this further simplifies to
\be
\begin{split}
\CK^{ab}\approx &\,
\delta^{ab}+ f^a_b+f_b^a + f^a_i f^b_i+(\tH\tH)^{ab}
\\
&\,
+\delta^{ab}(\tH^i_i)^2+2(\tH^a_b+\tH^b_a)\tH^i_i+\eps^{ajk}\tH^k_j\eps^{bmn}\tH^n_m.
\end{split}
\ee
The expression in the first line gives the fluctuation of the metric defined as a symmetric combination
of the two metrics $g^\pm$ constructed from $B^\pm$ and expanded around a bi-flat background.
All terms in the second line contain contributions either from trace of $\tH$ or from its antisymmetric part.
These components are constrained by the constraints $\hCH$ and $\hCD$, respectively.
However, they are restricted not to vanish, but to be given (in our approximation) by derivatives of $\eta$:
\be
\tH^i_i=\frac{1}{2\alpha}\, \eps^{ijk}\p_i\eta_j^k,
\qquad
\eps_{ajk}\tH^k_j=-\frac{1}{2\alpha}\,\p_i\eta_a^i.
\ee
Thus, we found that although in the linear approximation the fluctuations of the diagonal triad $\tE^a_i$ are consistent
with the fluctuations of the physical metric, already at the quadratic order the metric gets contributions from all fields.

\providecommand{\href}[2]{#2}\begingroup\raggedright\endgroup


\end{document}